\documentclass[3p,preprint]{elsarticle}

\journal{Theoretical Computer Science}

\usepackage{amssymb}
\newtheorem{theorem}{Theorem}
\newtheorem{lemma}[theorem]{Lemma}
\newtheorem{proposition}[theorem]{Proposition}
\newtheorem{corollary}[theorem]{Corollary}
\newtheorem{definition}[theorem]{Definition}

\newtheorem{example}{Example}
\newtheorem{remark}{Remark}

\newproof{proof}{Proof}

\usepackage[utf8]{inputenc}
\usepackage{complexity}
\usepackage[algosection]{algorithm2e}
\usepackage{endnotes}
\usepackage{listings}
\usepackage{mathtools}
\usepackage{verbatim}
\usepackage{mathrsfs}
\usepackage{longtable}
\usepackage{enumitem}
\usepackage{etoolbox}
\usepackage{float}
\usepackage{color}
\usepackage{hyperref}
\usepackage{multicol}

\usepackage{mathtools}

\newcommand*{\DEBUG}{}%

\ifdefined\DEBUG
\newcommand{\fixme}[1]{{\textcolor{red}{\bf{\textsf{FIXME: #1}}}}}
\newcommand{\bug}[1]{{\textcolor{blue}{\bf{\textsf{BUG: #1}}}}}
\newcommand{\idea}[1]{{\textcolor{blue}{\bf{\textsf{IDEA: #1}}}}}

\newcommand{\TODO}[1]{{\textcolor{red}{\bf{\textsf{
TODO: #1
}}}}}
\else
\newcommand{\fixme}[1]{}
\newcommand{\bug}[1]{}
\newcommand{\TODO}[1]{}
\newcommand{\idea}[1]{}
\fi

\newclass{\COMSLIP}{COM\mbox{-}SLIP}
\newclass{\COMSLIPCUP}{COM\mbox{-}SLIP^{\cup}}

\newclass{\DCM}{DCM}
\newclass{\eDCM}{eDCM}
\newclass{\eNPDA}{eNPDA}
\newclass{\CS}{CS}
\newclass{\TREE}{TREE}
\newclass{\DPDA}{DPDA}
\newclass{\RDPDA}{RDPDA}
\newclass{\PDA}{PDA}
\newclass{\DCMNE}{DCM_{NE}}
\newclass{\TwoDCM}{2DCM}
\newclass{\NCM}{NCM}
\newclass{\eNCM}{eNCM}
\newclass{\eNQA}{eNQA}
\newclass{\eNSA}{eNSA}
\newclass{\eNPCM}{eNPCM}
\newclass{\eNQCM}{eNQCM}
\newclass{\eNSCM}{eNSCM}
\newclass{\DPCM}{DPCM}
\newclass{\NPCM}{NPCM}
\newclass{\NQCM}{NQCM}
\newclass{\NSCM}{NSCM}
\newclass{\NPDA}{NPDA}
\newclass{\TRE}{TRE}
\newclass{\NFA}{NFA}
\newclass{\DFA}{DFA}
\newclass{\NCA}{NCA}
\newclass{\DCA}{DCA}
\newclass{\DTM}{DTM}
\newclass{\NTM}{NTM}
\newclass{\NTMCM}{NTMCM}
\newclass{\DLOG}{DLOG}
\newclass{\CFG}{CFG}
\newclass{\ETOL}{ET0L}
\newclass{\EDTOL}{EDT0L}
\newclass{\CFP}{CFP}
\newclass{\ORDER}{O}
\newclass{\MATRIX}{M}
\newclass{\BD}{BD}
\newclass{\LB}{LB}
\newclass{\ALL}{ALL}
\newclass{\decLBD}{decLBD}
\newclass{\StLB}{StLB}
\newclass{\SBD}{SBD}
\newclass{\TCA}{TCA}
\newclass{\LL}{{\cal L}}
\newclass{\CFGS}{CFG\mbox{-}S}
\newclass{\RLGS}{RLG\mbox{-}S}
\newclass{\LGS}{LG\mbox{-}S}
\newclass{\RLG}{RLG}
\newclass{\LG}{LG}
\newclass{\CFGSC}{CFG\mbox{-}SC}
\newclass{\CFGMC}{CFG\mbox{-}MC}
\newclass{\LGMC}{LG\mbox{-}MC}
\newclass{\RLGMC}{RLG\mbox{-}MC}
\newclass{\CCFGS}{CCFG\mbox{-}S}
\newclass{\LCFGSC}{LCFG\mbox{-}SC}
\newclass{\CLGS}{CLG\mbox{-}S}
\newclass{\CRLGS}{CRLG\mbox{-}S}
\newclass{\LGSC}{LG\mbox{-}SC}
\newclass{\RLGSC}{RLG\mbox{-}SC}

\newclass{\CSA}{CSA}

\newclass{\DCSA}{DCSA}
\newclass{\NCSA}{NCSA}
\newclass{\GSM}{GSM}
\newclass{\RDCSA}{RDCSA}
\newclass{\RNCSA}{RNCSA}
\newclass{\DCSACM}{DCSACM}
\newclass{\NCSACM}{NCSACM}

\DeclareMathOperator{\alp}{alph}

\DeclareMathOperator{\comm}{comm}

\newcommand{\shuffle}{\hspace{1mm}{\mathbin{\mathchoice
{\rule{.3pt}{1ex}\rule{.3em}{.3pt}\rule{.3pt}{1ex}
\rule{.3em}{.3pt}\rule{.3pt}{1ex}}
{\rule{.3pt}{1ex}\rule{.3em}{.3pt}\rule{.3pt}{1ex}
\rule{.3em}{.3pt}\rule{.3pt}{1ex}}
{\rule{.2pt}{.7ex}\rule{.2em}{.2pt}\rule{.2pt}{.7ex}
\rule{.2em}{.2pt}\rule{.2pt}{.7ex}}
{\rule{.3pt}{1ex}\rule{.3em}{.3pt}\rule{.3pt}{1ex}
\rule{.3em}{.3pt}\rule{.3pt}{1ex}}\mkern2mu}}\hspace{1mm}}

\begin{document}

\begin{frontmatter}

\title{State Grammars with Stores
\tnoteref{t1}\tnoteref{t2}}

\tnotetext[t1]{\textcopyright 2022. This manuscript version is made available under the CC-BY-NC-ND 4.0 license \url{http://creativecommons.org/licenses/by-nc-nd/4.0/} The manuscript is published in O.H. Ibarra, I. McQuillan. State Grammars with Stores. {\it Theoretical Computer Science} 798, 23--39 (2019).}

\tnotetext[t2]{A preliminary version of this paper has appeared in the 
Springer LNCS Proceedings of
the 20th International Workshop on Descriptional Complexity of Formal Systems (DCFS 2018), pp. 163--174.}

\author[label1]{Oscar H. Ibarra\fnref{fn1}}
\address[label1]{Department of Computer Science\\ University of California, Santa Barbara, CA 93106, USA}
\ead[label1]{ibarra@cs.ucsb.edu}
\fntext[fn1]{Supported, in part, by
NSF Grant CCF-1117708 (Oscar H. Ibarra).}

\author[label2]{Ian McQuillan\fnref{fn2}}
\address[label2]{Department of Computer Science, University of Saskatchewan\\
Saskatoon, SK S7N 5A9, Canada}
\ead[label2]{mcquillan@cs.usask.ca}
%\cortext[corr]{Corresponding author}
\fntext[fn2]{Supported, in part, by Natural Sciences and Engineering Research Council of Canada Grant 2016-06172 (Ian McQuillan).}

\begin{abstract}
State grammars are context-free grammars where the productions have states associated with them, and a production can only be applied to a nonterminal if the current state matches the state in the production. 
%In the past, several derivation relations have been used with state grammars to characterize the context-free languages, the languages generated by matrix grammars, and the recursively enumerable languages. Here, a derivation relation that operates in a circular fashion is introduced which also is shown to characterize the recursively enumerable languages. 
Once states are added to grammars, it is natural to add various stores, similar to machine models. With such extensions, productions can only be applied if both the state and the value read from each store matches between the current sentential form and the production. Here, generative capacity results are presented for different derivation modes, with and without additional stores. In particular, with the standard derivation relation, it is shown that adding reversal-bounded counters does not increase the capacity, and states are enough. Also, state grammars with reversal-bounded counters that operate using leftmost derivations are shown to coincide with languages accepted by one-way machines with a pushdown and reversal-bounded counters, and these are surprisingly shown to be strictly weaker than state grammars with the standard derivation relation
(and no counters). 
The complexity of the emptiness problem involving state
grammars with reversal-bounded counters is also studied.
%with an application: two results are provided concerning
%the succinctness of description of these grammars, and they are reduced to
%the question of whether $\P = \NP$.  

\end{abstract}


\end{frontmatter}

\section{Introduction} \label{sec:intro}

State grammars were created by Kasai \cite{Kasai}, and they have context-free grammar rules with additional state components. As originally defined, they consist of a set of nonterminals $V$, a set of terminals $\Sigma$, an initial nonterminal $S \in V$, a set of states $Q$, an initial state $q_0 \in Q$, and a set of productions $P$. Instead of normal context-free productions of the form $A \rightarrow w$, where $A \in V, w\in (V\cup \Sigma)^*$, now productions are of the form $(q,A) \rightarrow (p,w)$, where $q,p \in Q$, and $w$ was forced to be non-empty in Kasai's original formulation. Sentential forms are of the form $(q, \alpha)$ where $q \in Q, \alpha \in (V\cup \Sigma)^*$. A production is only applicable to a sentential form if the state of the production matches the state of the sentential form. The original derivation relation considered by Kasai (later called the {\em leftish} derivation relation in \cite{Moriya2005} which we will call it here as well), was as follows: $(q, u A v) \Rightarrow_{\rm lt} (p, u w v)$ if $(q,A) \rightarrow (p,w) \in P$, and $A$ is the leftmost nonterminal in the sentential form that has a production that is applicable from the current state. A word is generated if there is some leftish derivation starting at the initial state  and initial nonterminal that produces a word over $\Sigma^*$. The family of languages generated by such systems with $\lambda$-free rules, denoted by $\LL_{\rm lt}(\lambda\mbox{-free-}\CFGS)$, was shown to be equal to the family of context-sensitive languages \cite{Kasai}. Later, it was shown that when including $\lambda$ rules, the family produced, $\LL_{\rm lt}(\CFGS)$, is equal to the family of recursively enumerable languages \cite{Salomaa1972}. 

The definition of state grammars was extended shortly afterwards by Moriya \cite{Moriya1973} to also include a final state set $F$. Furthermore, he defined another derivation relation called the {\em free interpretation}, whereby any nonterminal can be rewritten that has a production defined on the current state, rather than the leftmost. With this derivation relation, the family of languages generated by state grammars, $\LL(\CFGS)$, was proven to equal the languages generated by matrix grammars (or $\lambda$-free matrix grammars for $\lambda$-free state grammars) \cite{DP1989}.

The notion of combining grammars with states is a powerful one. It becomes easy and natural to add various stores to grammars that operate like machine models. It can also enable the study of trade-offs between numbers of states, nonterminals, productions, and stores, relevant to the area of descriptional complexity. Changing the derivation relation and the rules allowed can also significantly change the families generated, obtaining many important language families as special cases. 

In this paper, we will collate some of the existing generative capacity results on state grammars. In doing so, we provide a shorter alternative proof that state grammars (with the free interpretation) generate the same family as matrix grammars by using context-free grammars with regular control. A new derivation mode is defined where all nonterminals are rewritten from left-to-right until the last nonterminal, then this repeats starting again at the first nonterminal. State grammars with this mode are found to generate the recursively enumerable languages (or context-sensitive languages for $\lambda$-free grammars). We will then consider adding multiple reversal-bounded counters to state grammars (with the free interpretation) and find that this does not change the capacity beyond only having states. However, this system provides quite an easy way of describing languages. Furthermore, it is shown that leftmost derivations for state grammars are strictly weaker than leftmost derivations for state grammars with counters, which are then strictly weaker than state grammars with no counters using the free interpretation. Lastly, the complexity of the emptiness problem
for several restrictions of state grammars with counters %. and applications to
%the succinctness of description of these grammars 
is investigated.

% Many proofs are omitted due to space constraints.

\section{Preliminaries}

%Here, some notation used in the paper is presented; 
We refer to \cite{HU} for an introductory treatment of automata and formal languages. We assume knowledge of deterministic and nondeterministic finite automata, context-free grammars, context-sensitive languages, and the recursively enumerable languages.

An {\em alphabet} $\Sigma$ is a finite set of symbols, a {\em word} over $\Sigma$ is a finite sequence of symbols
$a_1 \cdots a_n$, $n \geq 0, a_i \in \Sigma, 1 \leq i \leq n$, and $\Sigma^*$ (respectively $\Sigma^+$) is the set of all words (non-empty words) over $\Sigma$. The set $\Sigma^*$ contains the empty word, denoted by $\lambda$. Given a word $w \in \Sigma^*$, the length of $w$ is denoted by $|w|$, for $a \in \Sigma$, $|w|_a$ is the number of $a$'s in $w$, and
for subsets $X$ of $\Sigma$, $|w|_X = \sum_{a \in X} |w|_a$. The set of letters occurring in $w$ is denoted by $\alp(w) = \{a \in \Sigma \mid |w|_a >0\}$, and for $X \subseteq \Sigma$, $\alp_X(w) = \{a \in X \mid |w|_a >0\}$.
Given $\Sigma = \{a_1, \ldots, a_k\}$, the Parikh image of $w$ is $\psi(w) = ( |w|_{a_1} , \ldots, |w|_{a_k})$, extended to languages $L$, $\psi(L) = \{\psi(w) \mid w \in L\}$.
The commutative closure of $L$ is denoted by $\comm(L) = \{v \in \Sigma^*  \mid \psi(v) = \psi(w)$ for some $w \in L\}$.
We will not define the notion of semilinear sets and languages here, but an equivalent definition is that a language $L$ is semilinear if and only if it has the same commutative closure as some regular language \cite{harrison1978}.
Given $u,v \in \Sigma^*$, the shuffle of $u$ and $v$, denoted by $u \shuffle v$ is $\{u_1 v_1 \cdots u_n v_n \mid u = u_1 u_2 \cdots u_n, v = v_1 v_2 \cdots v_n, u_i,v_i \in \Sigma^*, 1 \leq i \leq n\}$.

The context-free languages are denoted by $\LL(\CFG)$, the linear languages are denoted
by $\LL(\LG)$, the context-sensitive languages by $\LL(\CS)$, and the right linear (regular languages) are denoted by $\LL(\REG)$.

Moreover, we will discuss other families and grammars systems summarized in \cite{DP1989}, such as matrix grammars. The languages generated by matrix grammars are denoted by $\LL(\M)$, and the languages generated by $\lambda$-free matrix grammars are denoted by $\LL(\lambda\mbox{-free-}\M)$.

\section{State Grammars}
 
We will formally define state grammars with final (accepting) states,
following the notation of \cite{Moriya1973}.

\begin{definition}
A {\em state grammar} ($\CFGS$), is a $7$-tuple $G = (V,\Sigma,P,S, Q,q_0,F)$,
where $V$ is the finite nonterminal alphabet, $\Sigma$ is the finite terminal
alphabet, $S \in V$ is the initial nonterminal, $Q$ is the finite set of states ($V,\Sigma,Q$ are pairwise disjoint), $q_0 \in Q$ is the initial state, $F\subseteq Q$ is the set of final states, and $P$ is a finite set of productions of the form:
$$(q,A) \rightarrow (p, w),$$
where $A \in V, w \in (V \cup \Sigma)^*, q,p \in Q$. 
The grammar is said to be linear (called $\LGS$) if, for all productions $(q,A) \rightarrow (p, w),$  $w \in \Sigma^* (V \cup \{\lambda\}) \Sigma^*$. 
The grammar is said to be right linear (called $\RLGS$) if, for all productions $(q,A) \rightarrow (p, w),$ $w \in \Sigma^* (V \cup \{\lambda\})$. 
In all cases, $G$ is $\lambda$-free if all productions are to some $(p,w)$ where $w \in (V \cup \Sigma)^+$.

A {\em sentential form} of $G$ is any element of $Q \times (V \cup \Sigma)^*$. 
Four different methods of derivation will be defined, with the last one being new. 
They are as follows:
\begin{enumerate}
\item The {\em free interpretation} derivation relation is defined such that $(q,u A v) \Rightarrow (p,u x v)$ if,  $(q,A) \rightarrow (p,x) \in P$, and $u,v \in (V \cup \Sigma)^*$. This is extended to the reflexive, transitive closure $\Rightarrow^*$. The language generated by $G$ is 
$$L(G) = \{w \mid (q_0,S) \Rightarrow^* (f,w), f \in F, w \in \Sigma^*\}.$$

\item The {\em leftmost derivation} relation is defined such that 
$(q,u A v) \Rightarrow_{\rm lm} (p,u x v)$ if, $(q,A) \rightarrow(p,x) \in P$, $(q,uA) \Rightarrow (p,ux)$, and $u \in \Sigma^*$. This is extended to the reflexive, transitive closure $\Rightarrow_{\rm lm}^*$.
The leftmost language generated by $G$ is 
$$L_{\rm lm}(G) = \{w \mid (q_0,S) \Rightarrow_{\rm lm}^* (f,w), f \in F, w \in \Sigma^*\}.$$

\item The {\em leftish} derivation relation is defined such that 
$(q,u A v) \Rightarrow_{\rm lt} (p,u x v)$ if, $(q,A) \rightarrow (p,x) \in P$, $(q,uA) \Rightarrow (p,ux)$, and for all $B \in \alp(u)$ with $B \in V$, then there is no production from $(q,B)$. This is extended to the reflexive, transitive closure $\Rightarrow_{\rm lt}^*$.
The leftish language generated by $G$ is 
$$L_{\rm lt}(G) = \{w \mid (q_0,S) \Rightarrow_{\rm lt}^* (f,w), f \in F, w \in \Sigma^*\}.$$

\item The {\em circular} derivation relation is,
for $v_0 A_1 v_1 \cdots A_n v_n, A_i \in V, v_i \in \Sigma^*$, $ 0 \leq i \leq n$,
\begin{align*}
&(p_0, v_0 A_1 v_1 A_2 \cdots A_n v_n) \Rightarrow_{\circ} (p_1, v_0 x_1 v_1 A_2 \cdots A_n v_n) \Rightarrow_{\circ} \\ & (p_2, v_0 x_1 v_1 x_2 v_2 A_3 \cdots A_n v_n) \Rightarrow_{\circ} \cdots \Rightarrow_{\circ} 
(p_n, v_0 x_1 v_1 x_2 v_2 \cdots x_n v_n),
\end{align*}
where $(p_i, A_{i+1}) \rightarrow (p_{i+1}, x_{i+1}) \in P$ for all $i$, $0 \leq i <n$. In this case, it is written
$$(p_0, v_0 A_1 v_1 A_2 \cdots A_n v_n) \Rightarrow_{\bullet} (p_n, v_0 x_1 v_1 x_2 v_2 \cdots x_n v_n).$$

This is extended to $\Rightarrow_{\bullet}^*$, the reflexive, transitive closure of $\Rightarrow_{\bullet}$. Therefore, this relation rewrites all nonterminals from left-to-right, then repeats in a circular fashion. The circular language generated by $G$ is
$$L_{\bullet}(G) = \{w \mid (q_0,S) \Rightarrow_{\bullet}^* (f,w), f \in F, w \in \Sigma^*\}.$$
\end{enumerate}
We also sometimes associate a new alphabet $\hat{P}$ in bijective correspondence with the productions of $P$, and write $u \xRightarrow[p]{} v$ where $p \in \hat{P}$, if the production associated with $p$ was applied from $u$ to $v$ (and similarly for the other derivation relations). %We sometimes call the free interpretation, the standard derivation relation.

The family of languages generated by $\CFGS$ grammars with the free interpretation (respectively the leftmost, leftish, and circular) derivation relation is
denoted by $\LL(\CFGS)$ (respectively $\LL_{\rm lm}(\CFGS)$, $\LL_{\rm lt}(\CFGS)$, $\LL_{\bullet}(\CFGS)$). 
For each of these families, we precede the family with $\lambda$-free to represent those languages generated by $\lambda$-free systems; e.g. $\LL_{\bullet}(\lambda\mbox{-free-}\CFGS)$. Similarly, replacing $\CFGS$ with $\LGS$ in these (or $\RLGS$) restricts the families to grammars where the rules are linear (or right linear).
\end{definition}

\begin{example}
\label{CFGS}
Let $k \geq 2$, $\Sigma = \{a_1, b_1, \ldots, a_k,b_k\}$, and  $G_k = (V,\Sigma,P,S,Q,q_0,F)$ where
$Q = \{q_0,\ldots, q_{k-1}\}$, $F= \{q_0\}$, and $P$ contains:
\begin{itemize}
\item $(q_0,S) \rightarrow (q_0, A_1 A_2 \cdots A_k)$,
\item $(q_{i-1},A_i) \rightarrow (q_i,a_iA_ib_i) \mid (q_i,a_ib_i)$, for $1 \leq i <k$,
\item $(q_{k-1},A_k) \rightarrow (q_0,a_kA_kb_k) \mid (q_0,a_kb_k)$.
\end{itemize}
In any successful derivation using the free interpretation, states must follow a pattern in $q_0 (q_0 \cdots q_{k-1})^+ q_0$, and from
$q_{i-1}$, only productions on $A_i$ can be applied, and they all must terminate on the last pass. Hence, $L(G_k) = \{ a_1^n b_1^n \cdots a_k^n b_k^n \mid n>0\}$.
\end{example}

Our first result shows that the different derivation relations
for linear and right linear grammars with states are the same.
\begin{proposition}
\label{prop1}
%The following are true:
$~~~$
\begin{itemize}
\item 
$\LL(\LG) = \LL(\LGS) = \LL_{\rm lm}(\LGS)=\LL_{\rm lt}(\LGS)= \LL_{\bullet}(\LGS)$,
\item $\LL(\REG) = \LL(\RLGS)= \LL_{\rm lm}(\RLGS)=\LL_{\rm lt}(\RLGS)= \LL_{\bullet}(\RLGS)$.
\end{itemize}
\end{proposition}
\begin{proof}
It is obvious that
the method of derivation does not matter for linear and right linear grammars.

A linear grammar (resp., a right linear grammar) can easily
be simulated by such a grammar with one state.
The converse follows by creating nonterminals
in $V \times Q$. For all productions of the form
$(q,A) \rightarrow (p, u B v), q,p \in Q, A,B \in V, u,v \in \Sigma^*$,
create a normal production $(q,A) \rightarrow u (p,B) v$ (i.e.\ the
state stays on the nonterminal; and for all terminating productions
of the form $(q,A) \rightarrow (p, u), q,p \in Q, A \in V, u \in \Sigma^*$,
create $(q,A) \rightarrow u$ if and only if $p \in F$. It is clear that
the languages generated are the same.
\qed
\end{proof}

The following was mentioned in \cite{Moriya2005}, and it follows by considering the standard simulation of context-free grammars with pushdown automata \cite{HU}, but using the state of the pushdown to simulate the state of the state grammar.
\begin{proposition}
$\LL_{\rm lm}(\CFGS) =\LL(\CFG) $.
\label{leftmost}
\end{proposition}
\begin{comment}
\begin{proof}
	Clearly, $\LL(\CFG) \subseteq \LL_{\rm lm}(\CFGS)$ by simulating each grammar with one state, since leftmost derivations do not change the language generated by a context-free grammar \cite{HU}.

For the converse, consider a $\CFGS$ $G$. The standard construction for converting $\CFG$s to $\NPDA$s \cite{HU} simulates a leftmost derivation by applying corresponding transitions in exactly the same order as the grammar (using only one state if accepting by empty stack). Hence, it is possible to simulate leftmost derivations of a $\CFGS$ by modifying the construction to follow the state transitions of $G$. Hence, $\LL_{\rm lm}(\CFGS) \subseteq \LL(\CFG) $.
\end{proof}
\end{comment}

As proven in \cite{Moriya1973}, the family of languages generated by matrix grammars (respectively $\lambda$-free matrix grammars) is equal to the family generated by state grammars (respectively $\lambda$-free state grammars) with the free interpretation. An alternate, shorter proof will be demonstrated by showing the equivalence of state grammars to context-free grammars with regular control \cite{DP1989}. It is known that such grammars are equivalent to matrix grammars \cite{DP1989}.
\begin{proposition}
\label{equalmatrix}
$\LL(\CFGS) = \LL(\MATRIX)$, and $\LL(\lambda\mbox{-free-}\CFGS) = \LL(\lambda\mbox{-free-}\MATRIX)$.
\end{proposition}
\begin{proof}
Instead of using matrix grammars, we use the equivalent formulation of context-free grammars with regular control (the equivalence holds for both the $\lambda$-free case, and the normal case \cite{DP1989}). This is a grammar $G = (V,\Sigma,P,S,R)$, where 
$G'= (V,\Sigma,P,S)$ is a context-free grammar, and for production labels $\hat{P}$
in bijective correspondence with $P$, $R \subseteq \hat{P}^*$ is a regular language, and the language generated by $G$ is 
$$L(G) = \{w \in \Sigma^* \mid S \xRightarrow[p_1]{} w_1 \xRightarrow[p_2]{} w_2 \xRightarrow[p_3]{} \cdots \xRightarrow[p_n]{} w_n = w \in \Sigma^*, p_1 \cdots p_n \in R\}.$$

Let $G$ be such a grammar, and let $M = (Q, \hat{P}, \delta, q_0, F)$ be a $\DFA$ accepting $R$. We construct a  $\CFGS$ grammar $G' = (V,\Sigma,P',S,Q,q_0,F)$ as follows:
for $p \in \hat{P}$ associated with $A \rightarrow w \in P$ and $\delta(q,p) = q', q, q' \in Q$, construct a production
$(q,A) \rightarrow (q',w) \in P'$.

Let $w \in L(G)$. Then $w_0 = S \xRightarrow[p_1]{} w_1 \xRightarrow[p_2]{} \cdots \xRightarrow[p_n]{} w_n = w \in \Sigma^*$ with $p_1 \cdots p_n \in L(M)$. Let $q_0' = q_0$ and $q_i' = \delta(q_{i-1}',p_i)$ for $i$, $1 \leq i \leq n, q_n' \in F$. Hence, if $p_i$ is associated with $A_i \rightarrow \alpha_i$, then $(q_{i-1}',A_i) \rightarrow (q_i',\alpha_i)$ is a production in $P'$. Hence,
$(q_0',S) \Rightarrow (q_1',w_1) \Rightarrow \cdots \Rightarrow (q_n',w_n)$, and $w \in L(G')$.

Conversely, let $w \in L(G')$. Then, 
$(q_0',w_0) \xRightarrow[\bar{p_1}]{} (q_1',w_1) \xRightarrow[\bar{p_2}]{} \cdots \xRightarrow[\bar{p_n}]{} (q_n',w_n)$, $q_n' \in F, w_n = w, w_0=S, q_0' = q_0$, and $\bar{p_1}, \ldots, \bar{p_n} \in \hat{P'}$. Let $p_i \in \hat{P}$ be the production letter obtained from the production associated with $\bar{p_i}$, for $1 \leq i \leq n$ by removing the states. By the construction of $\bar{p_i}$, $q_i' = \delta(q_{i-1}',p_i)$. Hence, $p_1 \cdots p_n \in R$, and $w \in L(G)$.

Next, let $ G = (V,\Sigma,P,S,Q,q_0,F)$ be a $\CFGS$. Let $P'$ be obtained from the productions in $P$ by removing
the states.
%Let $P_{\Sigma}$ be labels in bijective correspondence with productions in $P$. 
For $p \in P$, let $p' \in P'$ be the production obtained by removing the state, and if $\bar{p} \in \hat{P}$ is
the corresponding letter, then $\bar{p}' \in \hat{P}'$ is the corresponding letter. %Similarly, for $p \in P_{\Sigma}$, let $\bar{p}$ be the label of the production with the states removed, and $\bar{P_{\Sigma}}$ is this set of labels. 
Let $M = (Q,\hat{P'},\delta, q_0,F)$ be an $\NFA$ such that $\delta$ is defined as follows: if $\bar{p}$ is associated with $(q,A) \rightarrow (q',w) \in P$, then $q' \in \delta(q,\bar{p}')$. Furthermore, let $G' = (V,\Sigma,P',S,L(M))$ be a $\CFG$ with regular control.

Let $w \in L(G)$. Then $(q_0',w_0) \xRightarrow[\bar{p_1}]{} (q_1',w_1) \xRightarrow[\bar{p_2}]{} \cdots \xRightarrow[\bar{p_n}]{} (q_n',w_n), q_n' \in F, w_n = w, q_0'=q_0$. Then $\bar{p_1}' \cdots \bar{p_n}' \in L(M)$ and $w \in L(G')$.

Let $w \in L(G')$. Then $S = w_0' \xRightarrow[\bar{p_1}']{}  \cdots \xRightarrow[\bar{p_n}']{} w_n' = w$, with $\bar{p_1}' \cdots \bar{p_n}' \in L(M)$. Let $q_0', \ldots, q_n'$ be such that $q_0' = q_0, q_i' \in\delta(q_{i-1}',\bar{p_i}')$ and $q_n' \in F$. So, if
$\bar{p_i}'$ is associated with $A_i \rightarrow \alpha_i$, then $(q_{i-1}',A_i) \rightarrow (q_i' , \alpha_i) \in P$, and 
$(q_0',w_0') \Rightarrow \cdots \Rightarrow (q_n', w_n')$, and $ w\in L(G)$.
\qed
\end{proof}

We will see next that when circular derivations are used,
$\CFGS$ grammars already generate all recursively enumerable languages. 
We use the notion of a complete derivation tree $t$ of a context-free grammar \cite{HU}, which is a tree where all nodes are labelled by either a nonterminal, a terminal, or $\lambda$, the root is labelled by the initial nonterminal, if a parent is labelled by $A$ and its children are labelled by $A_1, \ldots, A_k$ from left to right, then $A \rightarrow A_1 \cdots A_k$ is a production, if a node is labelled by $\lambda$, then it is the only child of its parent, and all leaves are labelled by terminals. The yield of a derivation tree, ${\rm yd}(t)$, is the sequence of terminals obtained via a preorder traversal. Given such a tree $t$, level $i$ is all nodes at distance $i$ from the root, and the level-$i$ word is the sequence of labels on the nodes of level $i$ concatenated together from left to right. It is known that the set of yields of complete derivation trees of a grammar is exactly the language generated by the grammar \cite{HU}.

We will show that state grammars with circular derivations
are equivalent to {\em tree controlled grammars} which are defined as follows. A tree controlled grammar is a tuple $G = (V,\Sigma,P,S,R)$, where $G' = (V,\Sigma,P,S)$ is a context-free grammar, and $R$ is a regular language over $V \cup \Sigma$. When considering context-free derivation trees in $G'$, a restriction on the trees is used as follows: the language generated by $G$, $L(G)$, is equal to 
$$\{ {\rm yd}(t) \mid \begin{array}[t]{l} t \mbox{~is a complete derivation tree of~} G', \mbox{~for all levels $i$ but the last},\\ \mbox{the level-$i$ word is in~} R\}.\end{array}$$
Let $\LL(\TREE)$ (respectively $\LL(\lambda\mbox{-free-}\TREE)$) be the family of languages generated by (respectively $\lambda$-free) tree controlled grammars. It is known that tree controlled grammars generate all recursively enumerable languages, and $\lambda$-free tree controlled grammars generate exactly the context-sensitive languages \cite{DP1989}. %The proof appears in the appendix.
\begin{proposition} 
\label{tree}
%%The following are true:
$~~~$
\begin{itemize}
\item $\LL_{\bullet}(\CFGS) = \LL_{\rm lt}(\CFGS) = \LL(\TREE) = \LL(\RE) $,
\item $\LL_{\bullet}(\lambda\mbox{-free-}\CFGS) = \LL_{\rm lt}(\lambda\mbox{-free-}\CFGS)  = \LL(\lambda\mbox{-free-}\TREE) = \LL(\CS) $.
\end{itemize}
\end{proposition}
\begin{proof}
Obviously, $\LL_{\bullet}(\CFGS) \subseteq \LL(\RE)$. Also, every $\lambda$-free
$\CFGS$ with a circular derivation can be simulated by a linear bounded automaton \cite{HU} that simulates the sentential forms in a left-to-right fashion, eventually accepting if it matches the input.

For the reverse containment, we first show the case for $\lambda$-free grammars.
Let $G= (V,\Sigma,P,S,R)$ be an arbitrary $\lambda$-free tree controlled grammar. It is clear that we can assume without
loss of generality that $S$ does not appear on the right hand side of any production. Before constructing  a $\CFGS$,
we will transform $G$ into another tree controlled grammar $G_1 = (V_1, \Sigma,P_1,S^{\$},R_1)$ which generates the same language but  generates all terminals on the last level of each tree, the initial nonterminal only appears on the root of every tree, and the rightmost nonterminal on every level but the last (and only those) is tagged with a $\$$ symbol on the superscript.
Let $\bar{V_1} = V \cup \{X_a, \bar{X_a} \mid a \in \Sigma\}$, let
$V_1^{\$} = \{A^{\$} \mid A \in \bar{V_1}\}$, and let $V_1 = \bar{V_1} \cup V_1^{\$}$.
Hence, two new nonterminals $X_a$ and $\bar{X_a}$, are associated with each terminal $a \in \Sigma$, and then for each of these nonterminals, plus the nonterminals of $G$, another nonterminal tagged by $\$$ is also created.
Let $h$ be a homomorphism from $(V \cup \Sigma)^*$ to $\bar{V_1}^*$ that replaces $a \in \Sigma$ with $X_a$, and fixes each letter of $V$.
Let $f$ be a function from $\bar{V_1}^*$ to $(\bar{V_1}^* V_1^{\$}) \cup \{\lambda\}$ that puts a $\$$ as superscript on the last letter. Clearly, $f$ can be defined by a generalized sequential machine \cite{HU}.
Let $R_1 = f (h(R \cup \Sigma^*) \shuffle \{\bar{X_a} \mid a \in \Sigma\}^*)$.
Thus, $R_1$ is regular since the regular languages are closed under homomorphism, union, shuffle, and mappings defined by generalized sequential machines \cite{HU}. Next, create $P_1$ from $P$ via the following steps:
\begin{enumerate} 
\item For all productions $A\rightarrow w \in P$, create $A\rightarrow h(w)$ and $A^{\$} \rightarrow f(h(w))$.
\item For all $a \in \Sigma$, create  $X_a \rightarrow \bar{X_a},\bar{X_a} \rightarrow \bar{X_a}, \bar{X_a} \rightarrow a, X_a^{\$} \rightarrow \bar{X_a}^{\$},\bar{X_a}^{\$} \rightarrow \bar{X_a}^{\$}, \bar{X_a}^{\$} \rightarrow a$ to $P_1$.
\end{enumerate}
Notice $R_1$ is over $V_1$ and does not contain any terminals.

Let $t$ be a complete derivation tree of $G$, where the level-$i$ word, for all $i$ but the last, $n$ say, is in $R$. Let $t'$ be the $n+2$-level tree obtained from $t$ by replacing all nodes labelled by $a \in \Sigma$ in any level $i$ with a subtree with root $X_a$ that has $\bar{X_a}$ for all levels between $i$ and $n+1$, and $a$ at level $n+2$, and let $t''$ be obtained from $t'$ by tagging the rightmost node of every level but the last with $\$$ on the superscript. From the construction of $P_1$, $t''$ is a (context-free) complete derivation tree with the same yield. And, for each level-$i$ word, $i < n$, is in $h(R)$ with some number of letters from $\{\bar{X_a} \mid a \in \Sigma\}$ in it (which are allowed by the shuffle), with the final nonterminal tagged by $\$$ (using the function $f$). 
The level-$n$ word is in $f(h(\Sigma^*) \shuffle \{\bar{X_a} \mid a \in \Sigma\}^*)$, and the
level-($n+1$) word is in $f(\{\bar{X_a} \mid a \in \Sigma\}^*)$. Hence, $L(G) \subseteq L(G_1)$.

Let $t$ be a complete derivation tree of $G_1$, where the level-$i$ word for all $i$ but the last, $n$ say, is in $R_1$. For all such $i$, the last, and only the last nonterminal is tagged with $\$$. Without
the tag, each is in $h(R) \shuffle \{\bar{X_a} \mid a \in \Sigma\}^*$ or $h(\Sigma^*) \shuffle \{\bar{X_a} \mid a \in \Sigma\}^*$. Create a new derivation tree $t'$ as follows: for all subtrees rooted by 
$X_a$, they must have one child at every level, labelled by $\bar{X_a}$, until the last level, labelled by $a$; replace this subtree with a single node labelled by $a$. After this process, any level $i$ of $t$ with the level-$i$ word in $h(R) \shuffle \{\bar{X_a} \mid a \in \Sigma\}^*$ must correspond to a level-$i$ word in $t'$ in $R$. For any level $i$ of $t$ with the level-$i$ word $\alpha$ in 
$h(\Sigma^+) \shuffle \{\bar{X_a} \mid a \in \Sigma\}^*$, level $i$ would only have terminals in $t'$. If $\alpha'$ is obtained from $\alpha$ by removing letters of $\{\bar{X_a} \mid a \in \Sigma\}$, and changing $X_a$ to $a$, for $a \in \Sigma$, then the level-$i$ word in $t'$ would be $\alpha'$, and would
be labelled by $h^{-1}(\alpha)$, and would therefore be the last level and is ignored (i.e.\ the last
level is not verified to be in the regular language for a tree controlled grammar). All levels $i$ below that one
in $t$, have a level-$i$ word in $\lambda \shuffle \{\bar{X_a} \mid a \in \Sigma\}^*$, which do not exist in $t'$.
Thus,
the yield of $t$ is equal to that of $t'$, $t'$ is a complete derivation tree, and every level-$i$ word
before the last $i$ is in $R$. Hence, $L(G_1) \subseteq L(G)$.

Hence, $L(G) = L(G_1)$, and in $G_1$, all terminal derivations occur at the final level of the trees, the initial nonterminal $S^{\$}$ appears only at the root, and all complete derivation trees of $G_1$ have the rightmost node of every level except the last tagged with $\$$.

Let $M = (Q_1,V_1,\delta, q_0,F)$ be a $\DFA$ accepting $R_1$. Since every word ends with a symbol tagged with $\$$, we can assume without loss of generality that $F= \{q_f\}$, there are no transitions out of $q_f$, all transitions into $q_f$ are on a $\$$ tagged symbol, and there are no transitions into $q_0$. Let $G_2 = (V_1, \Sigma,P_2,S^{\$},Q_2, q_0,F_2)$ be a $\CFGS$, $Q_2 = Q_1 \cup \{q' \mid q \in Q_1\}$ and $F_2 = \{q_f'\}$,
with $P_2$ defined as follows: 
\begin{enumerate}
\item if $A\rightarrow \alpha \in P_1, A \in \bar{V_1}$ (i.e.\ not tagged with $\$$), $\alpha \notin \Sigma^*$, and $\delta(q,A) = p$ (and so $p \neq q_f$), create 
$(q,A) \rightarrow (p,\alpha)$,
\item if $A^{\$} \rightarrow \alpha \in P_1, A \in \bar{V_1}, \alpha \notin \Sigma^*$, and $\delta(q,A^{\$}) = q_f$, create both $(q,A^{\$})\rightarrow (q_0,\alpha)$ and $(q,A^{\$}) \rightarrow (q_0',\alpha)$,
\item if $A\rightarrow \alpha \in P_1, A \in \bar{V_1}$ (not tagged with $\$$), $\alpha \in \Sigma^*$, and $\delta(q,A) = p$ (and so $p \neq q_f$), create
$(q',A) \rightarrow (p',\alpha)$,
\item if $A^{\$} \rightarrow \alpha \in P_1, A \in \bar{V_1}, \alpha \in \Sigma^*$, and $\delta(q,A^{\$}) = q_f$, create $(q',A^{\$}) \rightarrow (q_f',\alpha)$.
\end{enumerate}
Notice that since all words in $L(M)$ end in a $\$$ tagged symbol, then the only productions to state $q_f'$ in $G_2$ are of the form $(q',\bar{X_a}^{\$}) \rightarrow (q_f',\alpha), \alpha = a \in \Sigma$.

We will show $L(G_1) = L(G_2)$.
Let $w \in L(G_1)$ with a complete derivation tree $t$ with $n$ levels, where the level-$i$ word is $\gamma_i$, for $1 \leq i \leq n$. Each level less than $n$ is tagged with a single $\$$ at the end. We know $\gamma_1, \ldots, \gamma_{n-1} \in L(M)$.
Then $t$ can be simulated in $G_2$ by deriving each level, one production at a time using a circular derivation with rules created in step 1 according to $\delta$, until the last node of each level. In $M$, reading this nonterminal takes $M$ to $q_f$. But in $G_2$, as long as it is before the second last level, it switches to $q_0$ with the first rule of step 2, so that it can continue the simulation at the next level. At the second last level, $G_2$ switches to $q_0'$ with the second rule of step 2 and it simulates this entire level with primed states created in step 3 (every production applied in this level generating terminals), and then the last $\$$-tagged nonterminal can switch to $q_f'$ with rules of type 4. Thus, $G_2$ can derive the yield of $t$. Hence, $L(G_1) \subseteq L_{\bullet}(G_2)$.

Conversely, let $w \in L_{\bullet}(G_2)$. Then
$$(p_0,\gamma_0 ) \Rightarrow_{\circ} (p_1, \gamma_1) \Rightarrow_{\circ} \cdots \Rightarrow_{\circ} (p_n,\gamma_n),$$ $\gamma_0 = S^{\$}, p_0 = q_0, p_n = q_f', \gamma_n = w\in \Sigma^*.$
Let $i_0,\ldots, i_m$ be such that $0 = i_0 \leq \cdots \leq i_m <n$, are exactly those indices where $(p_{i_j},\gamma_{i_j})$ is having its first nonterminal rewritten in the circular derivation.
Consider a complete derivation tree $t$ corresponding to this derivation where the nonterminals rewritten between $i_0$ and $i_1-1$ are at level $0$ from left-to-right, between $i_1$ and $i_2-1$ are at level $1$, etc.\ until $i_m$ to $n$ are at level $m$. Notice that by the construction, all terminal productions must appear at the last level of $t$ by the primed states of $G_2$. Also, since $p_n$ is the only state in the derivation that is $q_f'$, all of $\gamma_1, \ldots, \gamma_{n-1}$ end in a symbol tagged with $\$$, and no other $\$$-tagged symbol appears in each. And, in every context-free derivation tree of $G_2$ where all terminals appear in the last level, a $\$$-tagged symbol always appears in the rightmost node of each level between the first and the second last. Hence, in $t$, applying the state transitions of $G_2$ to the nonterminals in each level but the last from left-to-right leads from $q_0$ back to $q_0$, which implies that the level would be accepting in $M$. Hence, $w \in L(G_1)$ and $L_{\bullet}(G_2) \subseteq L(G_1)$.

The case where $\lambda$ productions are allowed is similar, except, if $A \rightarrow \lambda$ is a production, then $A^{\$} \rightarrow X_{\lambda}^{\$}$ needs to be included so that the tagged nonterminal does not end before the last level.
\qed
\end{proof}

Combining Propositions \ref{prop1}, \ref{leftmost}, \ref{equalmatrix}, and \ref{tree}, the known matrix languages that are not context-free, and the known context-sensitive languages that are not matrix languages \cite{Matrix1994}, the following hierarchies are obtained:

\begin{corollary}
%The following are true:
$~~~$
\begin{itemize}
\item $\LL(\CFG) = \LL_{\rm lm}(\CFGS) \subsetneq \LL(\CFGS) = \LL(\MATRIX) \subsetneq \LL_{\bullet}(\CFGS) =\LL_{\rm lt}(\CFGS) =\LL(\RE)$,
\item $\LL(\lambda\mbox{-free-}\CFG) = \LL_{\rm lm}(\lambda\mbox{-free-}\CFGS) \subsetneq \LL(\lambda\mbox{-free-}\CFGS) = \LL(\lambda\mbox{-free-}\MATRIX) \subsetneq \LL_{\bullet}(\lambda\mbox{-free-}\CFGS)$

$=\LL_{\rm lt}(\lambda\mbox{-free-}\CFGS) =\LL(\CS)$.
\end{itemize}

\end{corollary}

\section{State Grammars with Stores}

Now that states are attached to grammars, it is quite natural to attach one or more stores as well, just like machine models. Then, store contents can be part of sentential forms just as states are with state grammars. For example, one could define context-free grammars with states plus a pushdown store. This would be represented with a tuple just like a $\CFGS$ but with an additional word over the pushdown alphabet $\Gamma$ and a bottom-of-pushdown marker $Z_0$. In particular, the productions would be of the form $(q,X,A) \rightarrow (p, \alpha,w)$, where $q,p$ are states, $A$ is a nonterminal, $w$ is over the nonterminal and terminal alphabets, $X$ is the topmost symbol of the pushdown, and $\alpha$ is the string to replace the topmost symbol of the pushdown.
Sentential forms are therefore in $Q \times \Gamma^+ \times (V \cup \Sigma)^*$, and the derivation relation is defined in the obvious way.
Here, we will attach multiple reversal-bounded counters as stores as they are defined with reversal-bounded counter machines \cite{Ibarra1978}.
Explained briefly, a one-way $k$-counter machine is an $\NFA$ with $k$ counters, each containing some non-negative integer, and the transition function can detect whether each counter is empty or not, and can increment, keep the same, or decrement each counter  by one. Such  a machine is $r$-reversal-bounded if the number of times each counter switches between non-decreasing and non-increasing is at most $r$. Then $\LL(\NCM)$ is the family of languages accepted by machines that are $r$-reversal-bounded $k$-counter machines, for some $k,r\geq 1$. This type of machine is able to accept relatively complex languages, while maintaining a polynomial time emptiness problem when there are a fixed number of $1$-reversal-bounded counters \cite{Gurari1981220}. They therefore provide an interesting store to attach to state grammars.

Since grammars with states using either circular or leftish derivations already generate all
recursively enumerable languages, we will not consider those derivation relations
with stores.

Denote the set of all context-free grammars with states and some number of reversal-bounded counters by $\CFGSC$, and the languages they generate with the free interpretation and the leftmost derivation modes by $\LL(\CFGSC)$ and $\LL_{\rm lm}(\CFGSC)$ respectively.
For each such grammar $G= (V,\Sigma,P,S,Q,q_0,F)$ with $k$ counters, productions are of the form
$(q,i_1,\ldots, i_k,A) \rightarrow (p,l_1,\ldots, l_k,w)$, where
$p,q \in Q, i_j \in \{0,1\}$ (a production with $i_j = 0$ is applied if and only if counter $j$ is $0$),
$l_j \in \{-1,0,+1\}$ (which changes the counter), $A \in V, w \in (V \cup \Sigma)^*$.

\begin{example}
Let $G= (V,\{a,b\},P,S, Q,q_0,\{q_f\})$ be a $\CFGSC$ with $2$ counters accepting $\{w \$ w \mid w \in \{a,b\}^*, |w|_a = |w|_b\geq 0\}$, where $P$ is as follows:
\begin{itemize}
\item $(q_0, 0,0, S) \rightarrow (q_0, 0,0,A_1 A_2)$,
\item $(q_0, i,j, A_1) \rightarrow (q_a, 1,0, a A_1) \mid (q_b, 0,1, b A_1)$, for $i,j \in \{0,1\}$,
\item $(q_a, i,j, A_2) \rightarrow (q_0, 0,0, a A_2)$, $(q_b, i,j, A_2) \rightarrow (q_0, 0,0, b A_2)$, for $i,j \in \{0,1\}$,
\item $(q_0,i,i,A_1) \rightarrow (q_1, 0,0,A_1)$, for $i \in \{0,1\}$,
\item $(q_1,1,1,A_1) \rightarrow (q_1, -1,-1,A_1)$,
\item $(q_1,0,0,A_2) \rightarrow (q_1,0,0,\lambda)$, $(q_1,0,0,A_1) \rightarrow (q_f,0,0,\$)$.
\end{itemize}
To start, $G$ switches to $(q_0, 0,0,A_1A_2)$. Then the derivation repeatedly guesses either that $A_1$ derives an $a$ or a $b$; if it guesses it derives an $a$, it switches to $q_a$ and increases the first counter, and then from $q_a$, only $A_2$ can be rewritten and it must derive an $a$ (similarly with the $b$ case using $q_b$ and the second counter). Therefore, $A_1$ derives some sequence of terminals $w$ and $A_2$ must derive the same sequence, and the first counter contains $|w|_a$ and the second contains $|w|_b$. At any point while in state $q_0$, $G$ can switch to $q_1$ which repeatedly decreases both counters in parallel until both are verified to be zero at the same time, at which point $A_2$ is erased and $A_1$ generates $\$$.
\end{example}

Before studying the generative capacity of $\CFGSC$,
we need a definition.
A $\CFGSC$ $G$ is in normal form if each counter makes exactly $1$ reversal (once they decrease, they can no longer increase),
and a terminal string is successfully generated when $G$ enters a unique
accepting state $f$ and all the counters are zero. 
%Moreover, at each step at most one counter is changed (i.e., $+1$ or $-1$). 
We also assume 
that the state remembers when each counter enters a decreasing mode. So,
e.g., when counter $i$ enters the decreasing mode, the state remembers
that from that point on, counter $i$ can no longer increase. When another
counter $j$ enters the decreasing mode, the state now remembers that 
counters $i$ and $j$ can no longer increase, etc.
%The proof appears in the appendix.
\begin{lemma}
Let $G$ be an $\CFGSC$. We can effectively construct a $\CFGSC$ $G'$ in normal form
such that $L(G) = L(G')$.
\end{lemma}
\begin{proof}
Let $G= (V,\Sigma,P,S,Q,q_0,F)$ with $k$ $r$-reversal-bounded counters. We will describe how to create $G_1=(V \cup \{X\},\Sigma,P_1,S,Q_1,q_0,F_1)$, $X$ a new symbol, such that one of the counters (assume without loss of generality that it is the first counter) that is $r$-reversal-bounded is replaced with $n = \lfloor r/2\rfloor +1$ counters (the first $n$ counters of $G'$ are the new counters, and $G'$ has $n+k-1$ counters) that are $1$-reversal-bounded and $L(G) = L(G_1)$. 

Intuitively, instead of using counter 1 of $G$, $G'$ will use the first counter until the second reversal, then it will empty the first counter into the second counter until it is empty, then simulate using the second counter until the fourth reversal, etc., until the last reversal.

To do this, let, for $0 \leq i \leq n$, $Q_0^{(i)} = \{p_0^{(i)} \mid p \in Q\}$ (used to simulate transitions where the first counter is zero; when $i=0$, they are used to simulate when the counter is $0$ before the first increase, and when $i>0$, they are used after the $2i-1$st reversal and the counter returns to zero before increasing again (if it does return to zero).
For $1 \leq i \leq n$, let $Q_{\uparrow}^{(i)} = \{p_{\uparrow}^{(i)} \mid p \in Q\}$ (used when counter is increasing before the $2i-1$st reversal), and
$Q_{\downarrow}^{(i)} = \{p_{\downarrow}^{(i)} \mid p \in Q\}$ (used during the next decreasing section). 
For $2 \leq i \leq n$, let $\bar{Q}_{\uparrow}^{(i)} = \{\bar{p}_{\uparrow}^{(i)} \mid p \in Q\}$ (used to empty each counter into the next counter). Let $Q_1$ be the union of all these sets, and let
$F_1 = \{p_{\downarrow}^{(i)},p_{\uparrow}^{(i)}, p_{0}^{(j)} \mid 0 \leq j \leq n, 1 \leq i \leq n, p \in F\}$.

For all productions defined on $p \in Q$ with counter $1$ being empty that do not change the counter, $(p,0,i_2, \ldots, i_k,A) \rightarrow (r,0,j_2, \ldots, j_k,w)$ $p,r \in Q, i_2, \ldots i_k \in \{0,1\}, j_2, \ldots, j_k \in \{-1,0,+1\}$, make
\begin{eqnarray}
&&\mbox{for~} 0 \leq i \leq n, (p_0^{(i)},0, \ldots, 0, i_2, \ldots, i_k,A) \rightarrow (r_0^{(i)}, 0, \ldots, 0, j_2, \ldots, j_k,w), \label{step1}\\
&&\mbox{for~} 1 \leq i \leq n, (p_{\downarrow}^{(i)},0, \ldots, 0, i_2, \ldots, i_k,A) \rightarrow (r_0^{(i)}, 0, \ldots, 0, j_2, \ldots, j_k,w). \label{step2}
\end{eqnarray} 
For all productions on $p \in Q$ with counter $1$ being empty that increases the counter, $(p,0,i_2, \ldots, i_k,A) \rightarrow (r,+1,j_2, \ldots, j_k,w)$, $p,r \in Q, i_2, \ldots i_k \in \{0,1\}, j_2, \ldots, j_k \in \{-1,0,+1\}$, 
make
\begin{eqnarray}
&&\mbox{for~} 0 \leq i < n, (p_0^{(i)},0, \ldots, 0, i_2, \ldots, i_k,A) \rightarrow (r_{\uparrow}^{(i+1)}, l_1, \ldots, l_n, j_2, \ldots, j_k,w), \label{step3}\\
&&\mbox{for~} 1 \leq i < n, (p_{\downarrow}^{(i)},0, \ldots, 0, i_2, \ldots, i_k,A) \rightarrow (r_{\uparrow}^{(i+1)}, l_1, \ldots, l_n, j_2, \ldots, j_k,w), \label{step4}
\end{eqnarray}
where $l_{i+1} = +1$
and $l_m = 0$ for $m \neq i+1$.
For all productions defined on the counter being positive that increases or keeps the same counter value, $(p,1,i_2, \ldots, i_k,A) \rightarrow (r,z,j_2, \ldots, j_k,w)$ $z \in \{0,+1\}, p,r \in Q, i_2, \ldots i_k \in \{0,1\}, j_2, \ldots, j_k \in \{-1,0,+1\}$, make
\begin{eqnarray}
&&\mbox{for~} 1 \leq i \leq n, (p_{\uparrow}^{(i)},x_1, \ldots, x_n, i_2, \ldots, i_k,A) \rightarrow (r_{\uparrow}^{(i)}, l_1, \ldots, l_n, j_2, \ldots, j_k,w), \label{step5}
\end{eqnarray} 
where $x_i = 1$ and $l_i = x_i + z$, and $x_m = l_m = 0$, for $m \neq i$.
For those defined on the counter being positive that either decrease or keep the same counter value (note that multiple productions can be created from the same production), 
$(p,1,i_2, \ldots, i_k,A) \rightarrow (r,z,j_2, \ldots, j_k,w)$ $z \in \{0,-1\}, p,r \in Q, i_2, \ldots i_k \in \{0,1\}, j_2, \ldots, j_k \in \{-1,0,+1\}$, make
\begin{eqnarray}
&&\mbox{for~} 1 \leq i \leq n, (p_{\downarrow}^{(i)},x_1, \ldots, x_n, i_2, \ldots, i_k,A) \rightarrow (r_{\downarrow}^{(i)}, l_1, \ldots, l_n, j_2, \ldots, j_k,w), \label{step6}\\
&\mbox{if~} z = -1,&\mbox{for~} 1 \leq i \leq n, (p_{\uparrow}^{(i)},x_1, \ldots, x_n, i_2, \ldots, i_k,A) \rightarrow (r_{\downarrow}^{(i)}, l_1, \ldots, l_n, j_2, \ldots, j_k,w), \label{step7}
\end{eqnarray} 
where $x_i = 1$ and $l_i = z$, and $x_m = l_m = 0$, for $m \neq i$.
For those defined on the counter being positive that increase, 
$(p,1,i_2, \ldots, i_k,A) \rightarrow (r,+1,j_2, \ldots, j_k,w)$, $p,r \in Q, i_2, \ldots i_k \in \{0,1\}, j_2, \ldots, j_k \in \{-1,0,+1\}$, also make
\begin{eqnarray}
&&\mbox{for~} 1 \leq i < n, (p_{\downarrow}^{(i)},x_1, \ldots, x_n, i_2, \ldots, i_k,A) \rightarrow (\bar{r}_{\uparrow}^{(i+1)}, l_1, \ldots, l_n, j_2, \ldots, j_k,Xw), \label{step8}
\end{eqnarray} 
where $x_i = +1, x_m = 0$ for $m \neq i$, and $l_{i+1} = +1, l_m = 0$ for $m \neq i+1$ (notice that the new nonterminal $X$ is used here, which is the nonterminal that gets rewritten as the $i$th counter empties into the $i+1$st counter).
Lastly, make for all $2 \leq i \leq n$, 
$\bar{r}_{\uparrow}^{(i)} \in \bar{Q}_{\uparrow}^{(i)}, i_2, \ldots, i_k \in \{0,+1\},$
\begin{eqnarray}
&&(\bar{r}_{\uparrow}^{(i)},x_1, \ldots, x_n, i_2, \ldots, i_k,X) \rightarrow (\bar{r}_{\uparrow}^{(i)}, l_1, \ldots, l_n, 0, \ldots, 0,X), \label{step9}
\end{eqnarray} 
where $x_{i-1}=1, x_i \in \{0,1\}, x_m = 0, m \notin \{i,i+1\}, l_{i-1} = -1, l_i = +1, l_m = 0$ for $m \notin\{i,i + 1\}$; and
make
\begin{eqnarray}
&&\mbox{for~} 2 \leq i \leq n, (\bar{r}_{\uparrow}^{(i)},x_1, \ldots, x_n, i_2, \ldots, i_k,X) \rightarrow (r_{\uparrow}^{(i)}, 0, \ldots, 0,\lambda), \label{step10}
\end{eqnarray} 
where $x_i =1, x_m = 0, m \neq i$.

Consider a derivation of $G$,
$$(p_0, y_{0,1}, \ldots, y_{0,k}, \alpha_0) \Rightarrow \cdots \Rightarrow 
(p_{\beta}, y_{\beta,1}, \ldots, y_{\beta,k}, \alpha_{\beta}),$$
$\beta \geq 0, p_l \in Q, y_{l,j} \geq 0, \alpha_l \in (V \cup \Sigma)^*, 0 \leq l \leq \beta, 1 \leq j \leq k$, with $p_0 = q_0, \alpha_0 = S, p_{\beta} \in F, \alpha_{\beta} \in \Sigma^*, y_{0,j} = 0, 1 \leq j \leq k$. 
We refer to the sentential form with $l$ as subscript as configuration $l$.
For each $l$, $0 \leq l \leq \beta$, let $r_l$ be $i$, $ 0 \leq i \leq n$, if counter one has increased
after decreasing $i$ times by configuration $l$ (it is $0$ before having increased), and let
$$d_l = \begin{cases}
0 & \mbox{if either~} l = 0 \mbox{~or~} y_{l-1,1} = y_{l,1} = 0,\\
\downarrow & \mbox{if~} y_{l-1,1}>0 \mbox{~and the last production that changed counter 1 by config.~} l \mbox{~decreased},\\
\uparrow & \mbox{if~} y_{l,1}>0 \mbox{~and the last production that changed counter 1 by config.~} l \mbox{~increased.} 
\end{cases}$$
We will prove by induction that for all $l$, $0 \leq l \leq \beta$, there is a derivation of $G_1$,
$$((p_0)_0^{(0)},0, \ldots, 0,\alpha_0)\Rightarrow^* ((p_l)_{d_l}^{(r_l)}, z_{l,1},\ldots, z_{l,n}, y_{l,2}, \ldots, y_{l,k}, \alpha_l),$$
where $r_l > 0$ implies $z_{l,r_l} = y_{l,1}$, and $z_{l,m} =0$ for $m \neq r_l$.
When $l = 0$, then $r_l = 0, d_l = 0, y_{0,j} = 0, 1 \leq j \leq k, z_{0,i} = 0, 0 \leq i \leq n$, and the base case follows. Assume that it is true for $l$, $ 0 \leq l < n$, and consider configuration $l+1$.

\noindent {\bf Case 1}: the production applied between $l$ and $l+1$ does not change counter one.
Assume that $y_{l,1} = 0$, and then $y_{l,1} = y_{l+1,1} = 0$.
Either $d_l = 0$ or $d_l = \downarrow$. Then a production was created in step (\ref{step1}) or
(\ref{step2}) that leads to 
$((p_{l+1})_{0}^{(r_l)}, 0,\ldots, 0, y_{l+1,2}, \ldots, y_{l+1,k}, \alpha_{l+1}),$ where $r_{l+1} = r_l$
and induction follows. Assume $y_{l,1} > 0$. Induction follows similarly with a production of type
(\ref{step5}) or (\ref{step6}).

\noindent {\bf Case 2}: the production applied between $l$ and $l+1$ increases counter one.
Assume $d_l = 0$. Then a production created in (\ref{step3}) is created, and induction follows.
Assume $d_l = \uparrow$. Then a production created in (\ref{step5}) is created, and induction follows.
Assume $d_l = \downarrow$. If $y_{l,1} = 0$, then a production created in (\ref{step4})
is created, and induction follows with $r_{l+1} = r_l + 1$. If $y_{l,1} > 0$, then examining the
production created in (\ref{step8}) allows to rewrite to
$(\bar{p}_{l+1 \uparrow}^{(r_l+1)}, z_1', \ldots, z_n', y_{l+1,2}, \ldots, y_{l+1,k}, X \alpha_{l+1})$,
where $r_{l+1} = r_l + 1, z_{r_{l+1}}' = 1, z_m' = z_{l,m}$ for all $m \neq r_{l+1}$ and then using the production
created in (\ref{step9}) and (\ref{step10}), this can get rewritten to
$((p_{l+1})_{\uparrow}^{(r_{l+1})}, z_1, \ldots, z_n, y_{l+1,2}, \ldots, y_{l+1,k}, \alpha_{l+1})$
where $z_{r_{l+1}} = y_{l+1,1} = y_{l,1} + 1$, and $z_m = 0$ for all $m \neq r_{l+1}$. Thus, the induction follows.

\noindent {\bf Case 3}: the production applied between $l$ and $l+1$ decreases counter one.
Necessarily $y_{l,1}>0$, and so either $d_l = \uparrow$ or $d_l = \downarrow$.
Assume $d_l = \downarrow$; then using a production created in (\ref{step6}), then induction follows.
Assume $d_l = \uparrow$; then using a production created in (\ref{step7}), the induction follows. 

Thus, $\alpha_{\beta} \in L(G_1)$ as well.

Consider a derivation of $G_1$
$$(p_0,z_{0,1}, \ldots, z_{0,n},y_{0,2}, \ldots, y_{0,k}, \alpha_0) \Rightarrow
\cdots \Rightarrow 
(p_{\beta},z_{\beta,1}, \ldots, z_{\beta,n},y_{\beta,2}, \ldots, y_{\beta,k}, \alpha_{\beta}),$$
$\beta \geq 0, \alpha_{\beta} \in \Sigma^*, p_0 = (q_0)_0^{(0)}, z_{0,i} = 0, 1 \leq i \leq n, y_{0,j}=0, 2 \leq j \leq k, p_{\beta} \in F_1$. Each state $p_l$ is either of the form
$(t_l)_0^{(i)},(t_l)_{\uparrow}^{(i)} (t_l)_{\downarrow}^{(i)}$, or $\bar{(t_l)}_{\uparrow}^{(i)}$.
As the construction creates productions that change states as $G$ does (with the additional
subscripts and superscripts), and adds and subtracts to the first $n$ counters the same amount as $G$ does
to the first counter, then for all $l$ with $p_l$ of any of the first 3 forms above, there is a derivation of $G$,
$(q_0, 0, \ldots, 0, \alpha_0) \Rightarrow^* (t_l, z_{l,1} + \cdots + z_{l,n},y_{l,2}, \ldots, y_{l,k}, \alpha_l)$, and for the fourth form, the last component starts with $X$, which is removed in the derivation of $G$. Thus, $\alpha_{\beta} \in L(G)$.

Hence, $L(G) = L(G_1)$.

Applying this construction to each of the $k$ counters will create a machine with $kn$ $1$-reversal-bounded counters. Furthermore, it is evident that the grammar constructed is keeping track, using
the subscripts on the states, for each $1$-reversal-bounded counters, of which are empty, which are increasing and which are decreasing, as required by the normal form. From this grammar $G_1$ constructed with final
state set $F_1$, it is possible to build $G'$ in normal form by adding a new state $f$ that is the only
final state. Whenever $G'$ enters a state of $F_1$, $G'$ can nondeterministically switch to $f$
and then empty all counters. Hence, $G'$ is in normal form and generates the same language as $G$.
\qed
\end{proof}
Hence, we may assume that a $\CFGSC$ is in normal form.

Next, we will use this normal form to help show that reversal-bounded counters do not increase
the generative capacity.

\begin{proposition}
\label{removecounters}
$\LL(\CFGS) = \LL(\CFGSC) = \LL(\MATRIX)$.
\end{proposition}
\begin{proof}
By \cite{Moriya1973} (Proposition \ref{equalmatrix}), $\LL(\CFGS) = \LL(\MATRIX)$, and also it is immediate that $\LL(\CFGS) \subseteq \LL(\CFGSC)$.

Let $G = (V,\Sigma, P, S, Q, q_0,F)$ be a $\CFGSC$. Assume without loss 
of generality that $G$ is in normal form, and it therefore has $k$ $1$-reversal bounded counters.
Make a state grammar $G'$ (without counters) over $\Sigma \cup \Delta$ where $\Delta =  \{c_1, d_1, \cdots, c_k,d_k\}$ are new symbols. Then, $G'$ simulates $G$, but whenever it adds from counter $i$, it instead outputs terminal symbol $c_i$, and whenever it decreases from counter $i$, it outputs $d_i$. The states of $G'$ also verify that $G'$ starts by, for each counter $i$, 1) simulating only productions associated with counter $i$ being empty until it adds to the counter for the first time, 2) then it simulates productions defined on counter $i$ being positive (while outputting $c_i$'s), 3) then simulates productions on counter $i$ being positive (while outputting $d_i$'s), until some nondeterministically guessed spot after outputting some $d_i$, 4) where it then guesses that the counter is now empty, and then it only simulates productions on counter $i$ being empty while not outputting any more $c_i$'s and $d_i$'s. $G'$ operates in this fashion, as states were specifically marked in the normal form. Therefore, $G'$ operates just like $G$, where it simulates all of the counters, making sure that for each counter, all additions occur before any subtractions, but it does not do any of the counting. If one then restricts the derivations of $G'$ to those where the number of increases is the same as the number of decreases for each counter, then after erasing the letters of $\Delta$, it would give $L(G)$. Consider the following regular language $R = (c_1 d_1)^* \cdots (c_k d_k)^* \Sigma^*$, and the commutative closure of $R$, $\comm(R)$. Let $h$ be a homomorphism that erases all letters of $\Delta$ and fixes all letters of $\Sigma$. Then, $h(L(G') \cap \comm(R))$ is exactly this language, where $L(G') \cap \comm(R)$ restricts words to only those that have the same number of $c_i$'s as $d_i$'s (and hence the same number of increases as decreases for each counter) for each $i$, and $h$ erases the letters of $\Delta$. Hence, $L(G) = h(L(G') \cap \comm(R))$.

Since $G'$ is a normal state grammar, it can be converted to a matrix grammar $G''$ by Proposition \ref{equalmatrix}. It is known that the languages generated by matrix grammars are closed under intersection with $\LL(\NCM)$ \cite{matrix} (there, they used closure under the BLIND multicounter languages which is equivalent to $\LL(\NCM)$ \cite{G78}). Also, the commutative closure of every regular language is in $\LL(\NCM)$ \cite{eDCM}. So $L''' = L(G'') \cap \comm(R)$ is a language generated by a matrix grammar. Lastly, erasing all $c_i$'s and $d_i$'s with $h$ gives $L(G)$, and the languages generated by matrix grammars are closed under homomorphism \cite{DP1989}.  Since this gives a matrix grammar, it can be converted back to a normal state grammar (without counters) by Proposition \ref{equalmatrix} generating the same language as $L(G)$.
\qed
\end{proof}
We note that in the proof above, in $G$, despite the counters being $1$-reversal-bounded, productions can be applied to any nonterminal in the sentential form. Thus, some counter additions could occur when rewriting a nonterminal to the right of other nonterminals that get rewritten with a production that decreases. Hence, when intersecting with a language in $\LL(\NCM)$, it must not enforce that all $c_i$'s occur before any $d_i$'s.

The following corollary follows from the fact that the results
are true for matrix grammars \cite{Matrix1994}.

\begin{corollary}
%The following are true:
$~~~$
\begin{itemize}
\item Every unary language generated by a $\CFGSC$ is regular.
\item The emptiness problem for $\CFGSC$ is decidable.
\end{itemize}
\end{corollary}

Next, we will show that
$\LL_{\rm lm}(\CFGSC) = \LL(\NPCM)\subsetneq \LL(\CFGSC) $,
where $\NPCM$s are one-way nondeterministic pushdown automata augmented by reversal-bounded counters \cite{Ibarra1978}. Let $\NPCM(1)$ be $\NPCM$s where the pushdown is restricted to be $1$-reversal-bounded (once the pushdown pops, it can no longer push). We will need 
the notion of a $\CFG$ with monotonic counters introduced in
\cite{IbarraGrammars}. This is a simpler model of grammars with counters
that do not have states. At each step in the derivation,
the counters can be incremented by $0$ or $+1$, but not
decremented. A derivation in this grammar system starts with
the counters having value zero. A  terminal string $w$
is in the language of the grammar if there is a derivation
of $w$ that ends with all counters having the same value. 

Formally, a $\CFG$ with {\em monotonic counters} ($\CFGMC$) is a 5-tuple,
$G = (V, \Sigma,  P,  S)$, where
$\Sigma$ is the set of terminals,
$V$ is the set of nonterminals,
$S \in V$ is the initial nonterminal,
$k$ is the number of monotonic counters, all are initially set to 0, and
$P$ is the set of rules of the form: $A \rightarrow (c_1, \ldots, c_k, z)$,
where $A \in  V$, $c_i$ is either $0$ or $+1$, and $z \in (V \cup \Sigma)^*$.
The language defined is
$L(G)  = \{ w  ~|~  w  \in \Sigma^*,  (0, \ldots, 0, S)  \Rightarrow^* 
(n, \ldots, n, w)$ for some $n \ge 0 \}$. The languages generated are denoted by
$\LL(\CFGMC)$. Furthermore, $\LL_{\rm lm}(\CFGMC)$ are those generated by leftmost derivations.

\vskip .25cm

%\begin{comment}
%\noindent
%{\bf Example.}
\begin{example}
Consider the $\CFGMC$ $G$ over the 4-symbol alphabet $\{a, a', b, b'\}$
with two monotonic counters $C_1$ and $C_2$ and the
following rules:

\begin{itemize}
\item
$S \rightarrow (+1, 0, S a S a'S,) ~|~ (0, +1, S b S b' S) ~|~ (0, 0, \lambda)$.
\end{itemize}
\noindent
Clearly, the language we obtain when we ignore the counters in $G$
is the Dyck language $D_2$.  Then 
$L(G) = \{w ~|~ w \in D_2, |w|_a = |w_b| \}$.
%\end{comment}
\end{example}

If the rules of a $\CFGMC$ are of the form 
$A \rightarrow (c_1, \ldots, c_k, z)$, where $z \in
\Sigma^* V \Sigma^* \cup \Sigma^*$ (resp.,
in ($V \Sigma^* \cup \Sigma^*$)), then the grammar is called
an $\LGMC$ (resp., $\RLGMC$).
The following result was shown in \cite{IbarraGrammars}:

\begin{proposition}
$~~$
\begin{itemize}
\item
$\LL(\NPCM) = \LL(\CFGMC) = \LL_{\rm lm}(\CFGMC)$,
% is equal to the family of languages generated by $\CFGMC$s
\item
$\LL(\NPCM(1)) = \LL(\LGMC)$.
\item
$\LL(\NCM) = \LL(\RLGMC)$.
\end{itemize}
Moreover, the conversion between devices can be done in polynomial time.
\end{proposition}

From the proposition above, we have (where $\LGSC$ are linear state grammars with
counters, and $\RLGSC$ are right-linear state grammars with counters):

\begin{proposition}
\label{NPCMleftmost}
$~~~$
\begin{itemize}
\item
$\LL(\NPCM) = \LL(\CFGMC) = \LL_{\rm lm}(\CFGMC)=  \LL_{\rm lm}(\CFGSC)$.
\item
$\LL(\NPCM(1)) = \LL(\LGMC) = \LL(\LGSC)$.
\item
$\LL(\NCM) = \LL(\RLGMC) = \LL(\RLGSC)$.
\end{itemize}
Moreover, the conversion between devices can be done in polynomial time.
\end{proposition}
\begin{proof}
We will start with the first item.
Every $\CFGMC$ $G$ with a leftmost derivation can be simulated by a $\CFGSC$ with a leftmost derivation. It starts by simulating with one state. Then, before terminating, it guesses all counters are equal, decreases them all to zero to verify the counters were equal, then terminates. Thus, $\LL(\NPCM) \subseteq \LL_{\rm lm}(\CFGSC)$. For the reverse containment, consider the standard simulation of a $\CFG$ with a leftmost derivation by an $\NPDA$ \cite{HU}. This same construction can work with states while the counters of the $\CFGSC$ can be simulated by the counters of the $\NPCM$.

It is an easy construction to show that every $\LGMC$ (resp., $\RLGMC$)
can be simulated by an $\LGSC$ (resp., $\RLGSC$).  Also, following
the standard construction that simulates leftmost derivations of context-free grammars with $\NPDA$s \cite{HU}, linear grammars are simulated by a $1$-reversal-bounded pushdown automaton. By simulating the counters faithfully, every $\LGSC$ (resp., $\RLGSC$) can be simulated by an $\NPCM$ whose
stack is 1-reversal-bounded (resp., $\NCM$).
\qed
\end{proof}

\begin{proposition}
\label{NPCMmatrix}
$\LL(\NPCM) \subsetneq \LL(\MATRIX)$.
\end{proposition}
\begin{proof}
It is immediate that $\LL(\CFL) \subseteq \LL(\MATRIX)$, and it is known that $\LL(\MATRIX)$ is closed under intersection with $\LL(\NCM)$ \cite{matrix}, and homomorphism \cite{DP1989}. Recently, a Chomsky-Sch\"utzenberger-like theorem was shown that demonstrates that every language in $\LL(\NPCM)$ can be obtained by some Dyck language (which is context-free) intersected with an $\LL(\NCM)$ language, then mapped via a homomorphism \cite{IbarraGrammars}. Therefore, $\LL(\NPCM) \subseteq \LL(\MATRIX)$.

It is known that every $\NPCM$ language is semilinear \cite{Ibarra1978}. 
%Now $\LL(\CFGSC) = \LL(\MATRIX)$
%by Proposition \ref{equalmatrix}. 
It is known that matrix grammars can generate 
non-semilinear languages, e.g., a matrix grammar can generate
the non-semilinear language \cite{DP1989}, 
$L = \{a^n b^m  \mid  1  \le n  <  m  \le  2^n\}$.
It follows that $\LL(\NPCM) \subsetneq \LL(\MATRIX)$.
\qed
\end{proof}

\begin{comment}
\begin{proof} To show that $\NPCM$ is contained in $\CFGSC$, we just need
to show (by Proposition 2) that any language generated by a 
$\CFGMC$ can be generated by a $\CFGSC$. So let $G$ be a $\CFGMC$.  We 
construct a $\CFGSC$ with the following rules:
\begin{enumerate}
\item $(q_0, t_0, 0, \ldots, 0) \rightarrow (q_1, SX, 0, \ldots, 0)$ is a rule in $G'$,
where $q_0, q_1$ are states, $t_0$ is the (new) start nonterminal
of $G'$, $S$ is the start nonterminal of $G$, and $X$ is a new nonterminal.
\item If $A \rightarrow (z, c_1, \ldots, c_k)$ is a rule in $G$, then 
$(q_1, A, s_1, \ldots, s_k) \rightarrow (q_1, z, c_1, \ldots, c_k)$ is a rule in $G$,
where each $s_i$ is $0$ or $+$.
\item $(q_1, X, +, \ldots, +) \rightarrow (q_1, X, -1, \ldots, -1)$ is a rule in $G'$.
\item $(q_1, X, 0, \ldots, 0) \rightarrow (f, \lambda, 0, \ldots, 0)$ is a rule in $G'$.
\end{enumerate}

Clearly, $L(G') = L(G)$.

It is known that every $\NPCM$ language is semilinear (i.e., its 
Parikh map is a semilinear set). Now $\LL(\CFGSC) = \LL(\MATRIX)$
(Proposition 1). It known that Matrix grammars can generate 
non-semilinear languages, e.g., a matrix grammar can generate
the non-semilinear language, 
$L = \{a^n b^m  \mid  1  \le n  <  m  \le  2^n\}$.
It follows that $\NPCM$ is strictly contained in $\CFGSC$.
\qed
\end{proof}
\end{comment}

From Propositions \ref{leftmost}, \ref{equalmatrix}, \ref{tree}, \ref{removecounters}, \ref{NPCMleftmost}, and \ref{NPCMmatrix}, we have the
following hierarchy:

\begin{proposition} \label{prop13}
% The following is true:
$\LL(\CFG) = \LL_{\rm lm}(\CFGS) \subsetneq \LL_{\rm lm}(\CFGSC) = 
\LL(\NPCM) \subsetneq \LL(\CFGS) = 
\LL(\CFGSC) = \LL(\MATRIX) \subsetneq \LL_{\rm lt}(\CFGS) = \LL_{\bullet}(\CFGS) = \LL(\RE)$.
\end{proposition}
This demonstrates that state grammars with counters and leftmost derivations are strictly weaker than state grammars without counters.

\section{Controlled CFG-S}

We know that $\CFGS$ with leftmost derivations
are equivalent to $\CFG$ (Proposition \ref{leftmost}).
Under the free interpretation derivation,
$\CFGS$, $\CFGSC$, and $\M$ are equivalent (Proposition \ref{removecounters})
and these grammars have more generative power than 
$\CFGSC$ under leftmost derivation (Proposition \ref{prop13}).
We also know that $\NPCM$, $\CFGMC$, and $\CFGSC$ with
leftmost derivations are equivalent (Proposition \ref{NPCMleftmost}).
An interesting question is whether there is a natural restriction of state grammars without any
stores (no counters) that coincides with $\LL(\NPCM)$. In this section, such a restriction is provided.

Now define a controlled version of $\CFGS$, called $\CCFGS$ as follows:
A $\CFGS$ $G = (V,\Sigma,P,S,Q,q_0,F)$ is a $\CCFGS$ if
$V=V_1 \cup V_2$, where
$ S \in V_1$, $V_1$ is disjoint from $V_2 = \{C_1, \ldots, C_k\}$
for some $k \ge 1$, and the rules are of the
form $(q, A) \rightarrow (p,z)$, where 
\begin{itemize}
\item if $A \in V_1$, then  $z \in (V \cup \Sigma)^*$.
\item if $A  \in V_2$, then  $z \in V_2^*$. 
\end{itemize}
\noindent
We require the following concerning the application of the rules:
\begin{enumerate}
\item
If  $(q_, A) \rightarrow (p,z) \in P$ and $A\in V_1$, then this rule
is applicable to a sentential form if $A$ is the leftmost nonterminal
in $V_1$ that appears in the sentential form. (Note that nonterminals 
in $V_2$ can appear before $A$ in the sentential form.)
\item
A string $w \in L(G)$ if there exists a derivation
$(q_0, \alpha_0) \Rightarrow \cdots \Rightarrow (p_n, \alpha_n), p_0 = q_0, p_n \in F, \alpha_0 = S, \alpha_n \in \Sigma^*, w = \alpha_n$, such that
when a rule of the form $(q, C_i) \rightarrow (p,z)$ has been applied
where $z \in (V_2 - \{C_i\})^*$ ($1 \le i \le k$)
from sentential form $j-1$ to $j$ in the derivation,
then no more rules of the form $(q',X) \rightarrow (p', z')$
where $|z'|_{C_i} \ge 1$ can be applied from $j$ to $n$.
This means that once a $C_i$ is erased, 
no additional $C_i$'s can be added in the derivation.

\end{enumerate}
Note that strictly speaking, a new derivation relation is required that only rewrites the leftmost variable of $V_1$, but can rewrite any variable of $V_2$. However, we will only consider such a derivation relation informally. For the second condition,
the grammar itself can enforce it by adding states. So, 
if the state set of the grammar is $Q$, we can  expand the state set to 
$Q' = \{ [q, s_1,  \ldots,  s_k]  ~|~ q \in Q, s_i \in \{ +, -\} \}$. Then a rule of the form 
$(q, A) \rightarrow (p, z)$ is changed to: 
$([q, s_1, \ldots, s_k], A) \rightarrow ([p, s_1', \ldots, s_k'], z)$ 
where 
\begin{itemize}
\item Case:  $A \in V_1$: 
If $s_i = - $ then $s_i' = - $ and $z$ must not contain $C_i$; 
if $s_i = + $, then $s_i' = + $. 
\item Case:   $A = C_i$: 
If $s_i = - $, then  $s_i' = -$ and $z$ must not contain $C_i$ ; 
if $s_i = + $ and $z$ does not contain $C_i$, then $s_i' = - $; 
$s_j' = s_j$ for all $j \ne i$. 
\end{itemize}
The initial state is now $[q_0, +, \ldots, +]$ and the accepting 
states are now $[q, s_1, \ldots, s_k]$ where $q$ is accepting. 

\begin{comment}
Before proceeding further, we make the following observation:
We can generalize a one-way machine with 1-reversal-bounded counters
(e.g., NPCM, NCM, etc.) so that at every step, the machine 
can read a substring of the input (instead of only $\lambda$
or a single input symbol) and increment each counter by
any positive constant (instead of only by 1) and the resulting
machine is no more powerful than the original machine (by
using extra states). For the proofs below, we will use
the generalize machines.  
\end{comment}
    
 Even though this system does not have counters, it coincides with $\LL(\NPCM)$.
\begin{proposition} \label{prop14}
$\LL(\CCFGS) = \LL(\NPCM) = \LL_{\rm lm}(\CFGSC) = \LL(\CFGMC)$.
\end{proposition}
\begin{proof}
From Proposition \ref{NPCMleftmost}, we already know that
$\LL(\NPCM) = \LL_{\rm lm}(\CFGSC) = \LL(\CFGMC)$.

First, we show that $\LL(\CFGMC) \subseteq \LL(\CCFGS)$.
Let $G$ be a $\CFGMC$ with nonterminals $V$, start nonterminal $S$, and
$k$ counters $C_1, \ldots, C_k$. 
Construct a $\CCFGS$ $G'$ with states $q_0$ (start state), $q_1, \ldots, q_k, 
q_f$ (accepting state),
and $V = V_1 \cup V_2, V_1 = \{S'\} \cup V, V_2 = \{C_1,\ldots, C_k\}$, start nonterminal $S'$, and the following rules:
\begin{enumerate}
%\item
%$(q_0, S') \rightarrow (q_0, C_1 \cdots C_k S)$.
\item
$(q_0, S') \rightarrow (q_0, C_1 \cdots C_k S)$. 
\item
If $A \rightarrow (c_1, \ldots, c_k, z)$ is a rule in $G$
where $c_i \in \{ 0, 1\}$ ($1 \le i \le k$),
then add the following rule in $G'$:

$(q_0, A) \rightarrow (q_0, C_1^{c_1} \cdots C_k^{c_k} z) ~|~
(q_1, C_1^{c_1} \cdots C_k^{c_k} z)$.
\item
$(q_i, C_i) \rightarrow (q_{i+1}, \lambda)$ for $1 \le i < k$.

$(q_k, C_k) \rightarrow (q_1, \lambda) ~|~ (q_f, \lambda)$.
\end{enumerate}
\noindent
It is straightforward to verify that $L(G') = L(G)$ since the leftmost derivation of $G$ can be simulated
with only $q_0$, and at the last production of the simulation, it switches to $q_1$, where it
``decreases each counter'' one at a time (by erasing blocks of $C_1 C_2 \cdots C_k$ which do not have to be adjacent in the sentential form) before switching to a final state.

To complete the proof, we now show that $\LL(\CCFGS) \subseteq \LL(\NPCM)$.
Let $G = (V,\Sigma,P,S,Q,q_0,F)$ be a $\CCFGS$ with $V_2 = \{C_1, \ldots, C_k\}$. Without loss of generality,
assume that the accepting states of $G$
are halting, i.e., there are no transition rules from these states.
We construct an $\NPCM$ $M$ with $k$ 1-reversal-bounded counters $C_1, \ldots, C_k$,
initial stack symbol $Z$ (a new symbol), and $M$ has states consisting of the initial state $q_0$,
accepting state $q_f$ (a new state), $Q$, and some additional
states as needed in the simulation of $G$.
%Nov5 see above.

On input $w$, $M$ in state $q_0$ and top stack symbol $Z$,
first replaces $Z$ with $SZ$ while remaining in state $q_0$.  
(Convention: If $Y_1 \cdots Y_n$ is the
stack content, the leftmost symbol $Y_1$ is the
top of the stack, and $Y_n$ is the bottom of the stack.)
Then $M$ simulates the derivation of $w$ in $G$ just like
in the standard conversion of a $\CFG$ (with leftmost derivations) to an $\NPDA$ as follows:

\vskip .25cm

\noindent
Suppose that $M$ has just completed the simulation of a production rule
of $G$ and is in state $q$, and $q$ is not an accepting state of $G$.
To simulate a rule $(q,A) \rightarrow (p,z)$, $M$ executes the
following steps (using additional states):
%Nov5 above
\begin{enumerate}
\item
If $x$ is the terminal string to the left of the leftmost 
nonterminal symbol on the stack (note that $x  = \lambda$ if
the leftmost symbol is a nonterminal), then $M$ reads $x$
on the input while erasing $x$ from the stack.
\item 
If $A$ is in $V_1$ and $A$ is the top of the stack, then:
\begin{enumerate}
\item
$M$ increments counter $C_i$ by the number of $C_i$'s
in $z$ for $1 \le i \le k$.   
\item
$M$ rewrites $A$ by $y$, where $y$
is the string obtained from $z$ when the symbols in $V_2$
are deleted.
\item
$M$ enters state $p$.
\end{enumerate}
\item
If $A = C_i$ ($1 \le i \le k$), then for any top symbol in the stack:
\begin{enumerate}
\item
If $C_i$ appears in $z$ and counter $C_i > 0$, then $M$ increments
counter $C_i$ by $d_i-1$, where $d_i$ is the number of $C_i$'s in $z$.
If $C_i$ does not appear in $z$ and $C_i > 0$, $M$ decrements
counter $C_i$ by 1 and $M$ remembers that it can no longer simulate a production where $C_i$ is on the right hand side.
\item 
$M$ increments counter $C_j$ by the number of
$C_j$'s in $z$ for for all $j \ne i$.
\item
$M$ enters state $p$.
\end{enumerate}
%\item
%$M$ enters an accepting state when the stack only contains $Z$
%and $G$ is in accepting state.
\end{enumerate}

\noindent
Suppose that $M$ has just completed the simulation of a transition rule
of $G$ and is in state $q$, and $q$ is an accepting state of $G$.
(Note that by assumption, there is no transition rule from state $q$,
since it is an accepting state.)
%Nov5 above
Then $M$ checks that the stack is of the form $xZ$, where $x$
is a terminal string. To do this, $M$ verifies that the remaining input is
$x$ (by popping the stack), and then enters state $q_f$ (i.e., $M$ accepts).
%Nov5 two lines above

Clearly, $M$ accepts $L(G)$.
\qed
\end{proof}

\begin{remark}
In the definition of $\CCFGS$, rules of the form $(q, A) \rightarrow  (p,z)$,
where $A$ is in $V_2$, assume that $z$ is in $V_2^*$. We tried
to generalize the model to see if we can get the same
result (as in the proposition above) in two cases: 
\begin{enumerate}
\item Allow $z$ to be in $(V_2 \cup \Sigma)^*$,
\item Allow $z$ to be in $(V_1 \cup V_2)^*$.
\end{enumerate}
However, we were unable to do so, as the following discussion
explains. 

Consider the first case. Define a grammar $G$ with
states $q_0$ (start state), $q_1, q_2$ (accepting state),
$q_a, q_b, p_a, p_b$, $V_1 = \{S\}$, $V_2 = \{C_1, C_2\}$, $\Sigma = \{a,b\}$,
and the following rules:
\begin{multicols}{2}
\begin{itemize}
    \item $(q_0, S) \rightarrow (q_a, C_1 C_2) ~|~ (q_b, C_1 C_2)$
    \item $(q_a, C_1) \rightarrow (p_a, aC_1) ~|~ (q_1, \lambda)$
    \item $(q_b, C_1) \rightarrow (p_b, bC_1) ~|~ (q_1, \lambda)$
    \item $(q_1, C_2) \rightarrow (q_2, \lambda)$
    \item $(p_a, C_2) \rightarrow (q_a, aC_2) ~|~ (q_b,aC_2)$
    \item $(p_b, C_2) \rightarrow (q_a, bC_2) ~|~ (q_b,bC_2)$ 
\end{itemize}
\end{multicols}
It is evident that $L(G) = \{xx ~|~ x \in (a+b)^* \}$ which, we
conjecture, cannot be accepted by an $\NPCM$.  

For the second case, we modify the grammar $G$ above to
a grammar $G'$ as follows: Let $V_1 = \{S,A, B\}$.
% and
%$q_3$ be another state, which would now be the accepting state. 
The rules are obtained by replacing all $a$'s with $A$'s and $b$'s with $B$'s in $P$, and adding
\begin{itemize}
\item $(q_2, A) \rightarrow (q_2, a)$,
\item $(q_2, B) \rightarrow (q_2, b)$.
\end{itemize}
\begin{comment}
\begin{multicols}{2}
\begin{itemize}
\item $(q_0, S) \rightarrow (q_a, A C_1 B C_2) ~|~ (q_b, A C_1 B C_2)$
\item $(q_a, C_1) \rightarrow (p_a, AC_1) ~|~ (q_1, \lambda)$
\item $(q_b, C_1) \rightarrow (p_b, BC_1) ~|~ (q_1, \lambda)$
\item $(q_1, C_2) \rightarrow (q_2, \lambda)$
\item $(p_a, C_2) \rightarrow (q_a, AC_2) ~|~ (q_b,AC_2)$
\item $(p_b, C_2) \rightarrow (q_a, BC_2) ~|~ (q_b,BC_2)$ 
\item $(q_2, A) \rightarrow (q_2, a) ~|~ (q_3, \lambda)$
\item $(q_2, B) \rightarrow (q_2, b) ~|~ (q_3, \lambda)$
\item $(q_3, A) \rightarrow (q_3, \lambda)$
\item $(q_3, B) \rightarrow (q_3, \lambda)$
\end{itemize}
\end{multicols}
\end{comment}
Again, $L(G') = \{xx ~|~ x \in (a+b)^*\}$.
\end{remark}

We now look at special cases of $\CCFGS$ where the
rules of the form $(q, A) \rightarrow (p,z)$,
where $A$ is $V_1$ are restricted.
After deleting symbols from $V_2$, if $z$ can only be in ($\Sigma^* V_1 \Sigma^* \cup \Sigma^*$),
then call the grammar a $\CLGS$.  Similarly, after deleting, if
$z$ can only be in ($V_1\Sigma^* \cup \Sigma^*$), then call
the grammar a $\CRLGS$. 

\begin{proposition}
$~~~$
\begin{itemize}
\item
$\LL(\CLGS) = \LL(\NPCM(1)) =
  \LL(\LGMC) = \LL(\LGSC)$.
\item
$\LL(\CRLGS) = \LL(\NCM) = \LL(\RLGMC) = \LL(\RLGSC)$.
\end{itemize}
\end{proposition}
\begin{proof}
Again, from Proposition \ref{NPCMleftmost}, we already know that
$\LL(\NPCM(1)) = \LL(\LGMC) = \LL(\LGSC)$,
and $\LL(\NCM) = \LL(\RLGMC) = \LL(\RLGSC)$.

Certainly, the first part of the proof of Proposition \ref{prop14} 
applies directly to showing that $\LL(\LGMC) \subseteq \LL(\CLGS)$
and $\LL(\RLGMC) \subseteq \LL(\CRLGS)$. 

For the proof that $\LL(\CLGS) \subseteq \LL(\NPCM(1))$, the construction needs to be modified so that terminals do not get pushed and then immediately popped, causing the $\NPCM$ to not be
$1$-reversal-bounded. Thus, modify
the construction of $M$ in the second part of the
proof of  Proposition \ref{prop14} as follows:

\vskip .25cm

\noindent
Delete item (1) and replace item (2) with the following:
\begin{enumerate}
\item []
If $A$ is in $V_1$ and $A$ is the top of the stack, then:
\begin{enumerate}
\item
$M$ increments counter $C_i$ by the number of $C_i$'s
in $z$ for $1 \le i \le k$.   
\item
-- If $z$, when the symbols in $V_2$ are deleted,
%Nov5 above
results in a string of the form $x_1 B x_2$ for some terminal
strings $x_1, x_2$ and nonterminal $B$, then $M$ reads
$x_1$ on the input and replaces $A$ with $Bx_2$. 

-- If $z$, when the symbols in $V_2$ are deleted,
%Nov5 above
results in terminal string $x$, $M$ reads $x$ on the input
and pops $A$ from the stack.
\item
$M$ enters state $p$.
\end{enumerate}
\end{enumerate}
\noindent
The rest of the construction is the same.

For the proof that $\LL(\CRLGS) \subseteq \LL(\NCM)$,
the construction above can trivially be modified, noting
that we do not need the stack; hence, $M$ is an $\NCM$.
\qed
\end{proof}

%Now $\LGS$ is equivalent
%to $\LG$, and $\RLGS$ is equivalent to $\RLG$
%(Proposition \ref{2}).

\section{Complexity of the Emptiness Problem}

In this section, we study the emptiness problem for restrictions of $\CFGSC$s.
Given a $\CFGS$ $G = (V,\Sigma,P,S,Q,q_0,F)$, a 
 derivation 
$(p_0,\alpha_0) \Rightarrow (p_1,\alpha_1) \Rightarrow \cdots \Rightarrow 
(p_n,\alpha_n)$ with $p_0 = q_0, \alpha_0 = S, p_n \in F,\alpha_n = w \in \Sigma^*$, is of index $m$ if
$|\alpha_i|_V \leq m$, for all $0\leq i \leq n$. The grammar $G$ is of index $m$ if, for every $w \in L(G)$, there exists some derivation of $w$ that is of index $m$. If it is index $m$ for some $m$, then
it is said to be finite-index.
This property is more general than requiring that every derivation of a word
in the language is of index $m$, a notion that is called
uncontrolled index $m$, or uncontrolled finite-index. These notions are well-studied for different types of grammars \cite{RozenbergFiniteIndexGrammars}. For context-free grammars, finite-index is more general than uncontrolled finite-index, as uncontrolled finite-index grammars correspond to pushdown automata with a reversal-bounded pushdown \cite{GS}, which cannot accept languages such as 
$\{a^n b^n \mid n>0\}^*$ that can be generated by an index $2$ grammar \cite{RozenbergFiniteIndexGrammars}.
This notion can be defined similarly for $\CFGMC$s and $\CFGSC$s as well.

%Hence, 
%Notice that 
% every string in the 
%generated language has some derivation where each sentential form
%in the derivation has at most $m$ nonterminals.  Note that 
%the case $m = 1$ corresponds to a linear context-free grammar (LCFG).

%We state the following proposition whose proof will appear
%in the full version of this paper:

We begin with the following lemma.
\begin{lemma} \label{new}
Given a binary number $x = b_1 b_2 \cdots b_k$
(with $b_1$ being the least significant bit), we can construct a $\CFGMC$
$G$ with a monotonic counter $C$ such that
$G$ generates $\lambda$
with its monotonic counter containing the number
represented by $x$.  Moreover, $|G|$
(the size of the grammar) is polynomial in $k$, can be built in
polynomial time, and
$G$ has finite index that is in $O(k)$.  
\end{lemma}
\begin{proof}
Let $x = b_1 b_2 \cdots b_k$, where each $b_j$ is either 0 or 1.
Let $n$ be the number represented by $x$.
We construct a $\CFGMC$ $G$ with one monotonic counter and
nonterminals $\{S\} \cup \{[i,0] ~|~ 1 \le i \le k\}
\cup \{[i,1] ~|~ 1 \le i \le k \}$ with the following rules:
\begin{enumerate}
\item
$S \rightarrow (0, [1, b_1] [2, b_2] \cdots [k, b_k])$,
\item
$[i, 0] \rightarrow (0, \lambda)$, for $1 \le i \le  k$,
\item
$[i, 1] \rightarrow (+1, [1, 1] [2, 1] \cdots [i-1, 1] [i, 0])$,
for $1 < i \leq k$,
\item
$[1, 1] \rightarrow (+1, [1, 0])$,
\end{enumerate}

\noindent
As $G$ has $2k+1$ nonterminals and $2k+1$ productions, each with at most
$k$ letters on the right hand side, the size of $|G|$ is $O(k^2)$.

Next, by induction on $l$, $1 \leq l \leq k$, we will prove that 
$(0,[l,1]) \Rightarrow \cdots \Rightarrow (2^{l-1},\lambda)$, which is a derivation of index $l$.

For $l=1$, $(0,[1,1]) \Rightarrow (1,[1,0]) \Rightarrow (1,\lambda)$, which is of index $1$.

Assume it is true for $l, 1 \leq l <k$. Then $(0,[l+1,1]) \Rightarrow 
(1,[1,1][2,1] \cdots [l,1][l+1,0])$. The last term generates $0$, while
the rest generates $2^0 + \cdots + 2^{l-1}$ by the inductive hypothesis, plus $1$ is added, which is equal to $2^l$.
Generating using a leftmost derivation, it first
generates from $[1,1]$ (that part of the derivation taking index $1$, with the entire sentential form being $l+1$ index and ending with a sentential form with $l$ nonterminals), then
$[2,1]$ (taking index $2$, total $l+1$, and ending with $l-1$ nonterminals), then $[3,1]$ (taking
index $3$, total $l+1$), etc. with $[l,1]$ taking
index $l$ bringing the total index of this derivation to $l+1$. Thus, the statement is true, and $G$ has
index $k+1$.
\qed
\end{proof}

This can be used to study non-emptiness for $\CFGSC$ grammars.
\begin{proposition} \label{part2}
The non-emptiness problem for $\CFGSC$ with the
leftmost derivation mode is $\NP$-complete.
In fact, it is $\NP$-hard even when the grammar
has finite index, and there is only one 1-reversal-bounded counter.
(Note that when there is no counter, the problem is polynomial-time 
decidable \cite{HU}.)
\end{proposition}
\begin{proof}
The first part ($\NP$-completeness in general) follows from Proposition \ref{NPCMleftmost}
and the fact that the non-emptiness problem for $\CFGMC$ is $\NP$-complete
\cite{IbarraGrammars}.

For the second part, we use 
the subset-sum problem, which is $\NP$-hard \cite{garey1979computers}.
An instance $I$ of this problem is the following: 

{\em Given}: $k \ge 1$, and positive integers 
$x_1, \ldots, x_k, x_{k+1}$ represented in binary.

{\em Question}: Is there a  subset of
\{$x_1, \ldots,x_k\}$ that sums to $x_{k+1}$?

\noindent   
From Lemma \ref{new}, we can construct for each $1 \le i \le k+1$,
a $\CFGMC$ $G_i$ with start nonterminal $S_i$
and one monotonic counter which
generates $\lambda$
%(under leftmost derivation)
with counter value $n_i$ (the number represented by $x_i$).

Assume that the nonterminals used in $G_1, \ldots, G_k, G_{k+1}$ are pairwise distinct.
Let $S_1, \ldots, S_k, S_{k+1}$ be their start nonterminals. Let $A_1,\ldots, 
A_{k+1}, Z$ be new symbols.  We construct a $\CFGSC$ $G$ with 
a unary terminal alphabet $\{a\}$, states
$q_0$ (the start state), $q_1$, $q_f$ (the accepting state),
and $A_1$ the start nonterminal.
$G$ has the following rules, where $v$ denotes 0 or 1:
\begin{enumerate}
\item \label{t1}
$(q_0, v, A_i) \rightarrow (q_0, 0, S_i A_{i+1}) ~|~  (q_0, 0, A_{i+1})$  
for $1 \le i \le k$.
\item \label{t2}
If  $X \rightarrow (c, w)$  is a rule in one of the $\CFGMC$s $G_1,
\ldots, G_k$ where $c$ is either $0$ or $1$,
then $(q_0, v, X) \rightarrow (q_0, c, w)$ is a rule in $G$.
\item \label{t3} $(q_0,v,A_{k+1}) \rightarrow (q_1,0,S_{k+1}Z)$.
\item \label{t4}
If  $X \rightarrow (0, w)$  is a rule in $\CFGMC$ $G_{k+1}$,
then $(q_1, v,  X) \rightarrow (q_1, 0, w)$ is a rule in $G$.
\item \label{t5}
If  $X \rightarrow (1, w)$  is a rule in $\CFGMC$ $G_{k+1}$,
then $(q_1, 1,  X) \rightarrow (q_1, -1, w)$ is a rule in $G$.
\item \label{t6}
$(q_1, 0, Z) \rightarrow (q_f, 0, a)$ is a rule in $G$.
\end{enumerate}

\noindent
Rules of type \ref{t1} are used to pick some subset $Y$ of $G_1, \ldots, G_k$
in order to add the respective subset of $x_1,\ldots, x_k$ to the counter.
Then for each $G_i, 1 \leq i \leq k$ that is in $Y$, any number that can be
added to the counter in $G_i$ can be added to $G$ using rules of type 
\ref{t2}. The entire process thus far occurs entirely in a leftmost fashion using state $q_0$
only, starting with $G_1$, optionally adding $x_1$ to the counter while generating
$\lambda$, then doing the same with $G_2$, etc.\ with $G_k$. At this point, the sentential
form is $(q_0,x, A_{k+1})$ where $x$ is the sum of counters of $X$. $G$ then can (and has to)
switch to $q_1$ using the rule of type \ref{t3}. Rules of type
\ref{t4} and \ref{t5} are then used to decrease any counter value that could be added to $G_{k+1}$.
Once this completes and the nonterminals of $G_{k+1}$ are erased, if the counter is zero, then
$Y$ must represent a solution, and type \ref{t6} is used, which generates the
terminal $a$ if and only if $Y$ is a solution. 

Hence, $L_{\rm lm}(G)$ is non-empty if and only if the subset-sum problem 
has a solution, which is $\NP$-hard. The reduction also runs in polynomial time as
the grammars $G_1, \ldots, G_k$ can be built in polynomial time, and $|G|$ is linear  in 
$|G_1| + \cdots + |G_{k+1}|$.  Furthermore, $G$ only has one counter that is $1$-reversal-bounded,
and has finite index, as it simulates each $G_i$ from $1 \leq i \leq k+1$, each of which is finite index by
Lemma \ref{new}.
\qed
\end{proof}

Observe that in the $\NP$-hardness proof above, $G$ has only one 1-reversal-bounded counter, and
is finite-index, but the index is not fixed but instead grows linearly with the maximum of
$|x_1|, \ldots, |x_{k+1}|$ (in binary), by Lemma \ref{new}.
In the proof above, suppose instead that $G$ has index 1 but the number of 1-reversal-bounded
counters is not fixed.  We will show that the non-emptiness problem
is also $\NP$-hard.

First, for the purposes of the proof below, we note that we can generalize the reversal-bounded 
counters in a $\CFGSC$ by allowing increments
in the counters to be binary constants $c$, where $c \ge 0$,
but decrements are still restricted to 1.
There can be many such $c$'s used in the rules, e.g., a
grammar with three counters $C_1, C_2, C_3$ can have rules like:
$$(q, 0, 1, 1, A) \rightarrow (p, +3, 0, +2, \alpha).$$
The above means: If in state $q$ and the number in $C_1$ is $0$,
the number in $C_2$ is positive, and the number in $C_3$ is
positive,  then $C_1$ is incremented
by 3, $C_2$ is left unchanged, $C_3$ is incremented by 2, $A$ is
replaced by $\alpha$, and the state changes to $p$.
We call these generalized $\CFGSC$. Other grammatical models with
states and reversal-bounded counters can similarly
be generalized. 

We can convert a generalized $\CFGSC$ $G$ to an equivalent $\CFGSC$
$G'$ by noting that a constant $c$ can be computed with 1-reversal-bounded
unit counters (i.e., can only increment/decrement by 1) efficiently.
For suppose $c = b_1 b_2 \cdots b_k$, where each $b_i$ is 0 or 1
(with $b_1$ the least significant bit).
Then the number corresponding to $c$ 
is $n = b_12^0 + b_2 2^1 + \cdots + b_k 2^{k-1}$.  
To compute and store $n$ in a counter, we need to compute
the terms and add them up. Clearly, a term $2^s$
can be computed with 1-reversal-bounded unit counters
using ``recursive doubling''.  For example, if $2^i$ is in a
counter $C_1$,  $2^{i+1}$ can be computed in counter $C_2$
by adding 1 to $C_2$ twice for every decrement of 1 in $C_1$.
The number of 1-reversal-bounded unit counters needed to compute $n$ would be polynomial in $k$.  We also need
to introduce new temporary nonterminals to implement the
conversion. 
For example, to simulate production $t$: $(q,1,A) \rightarrow (p, +5, \alpha)$, then
$5 = 1 \cdot 2^0 + 1 \cdot 2^2$. Then $G'$ has the one main counter ($C_1$) plus two extra
counters (none for the $2^0$ term and two for the $2^2$ term) for this rule. Corresponding to this rule, instead $G'$ uses
\begin{multicols}{2}
\begin{enumerate}
    \item $(q, 1,0,0,A) \rightarrow (p,+1,+1,0,X_1^t)$
    \item $(p,1,1,0,X_1^t) \rightarrow (p,0,+1,0,X_2^t)$
    \item $(p,1,1,z,X_2^t) \rightarrow (p,0,-1,+1,X_3^t), z \in \{0,1\}$
    \item $(p,1,1,1,X_3^t) \rightarrow (p,0,0,+1,X_2^t)$
    \item $(p,1,0,1,X_2^t) \rightarrow (p,+1,0,-1,X_2^t)$
    \item $(p,1,0,0,X_2^t) \rightarrow (p,0,0,0,\alpha)$.
\end{enumerate}
\end{multicols}
The first two rules add the least significant digit to the main counter, and $2$ to the 
second counter. Then, the third and fourth rule run in a loop, and they empty counter
two while doubling its content into counter 3. At this point, counter three
can be emptied into $C_1$, the new $X_i^t$ nonterminals go to the original right hand side $\alpha$, and the simulation continues.
Note that all these rules do not alter the index as all the new rules have one nonterminal
on the right hand side until rewriting to $\alpha$.
 The number of rules and the number of counters would also be polynomial in
$|c|$.  Hence, $|G'|$ (the size of $G$) is
polynomial in $|G|$, and can be constructed in polynomial time.
Also, if $G$ has index $m$, then the $G'$ constructed as described
will also have index $m$. 
This can be summarized as follows:
\begin{lemma}
Given a generalized $\CFGSC$ $G$, a $\CFGSC$ $G'$ can be constructed in polynomial time such that
$|G'|$ is polynomial in $|G|$, the number of counters is at most $|G|$, and if $G$ is of index $m$,
then so is $G'$. Furthermore, if $G$ is $\RLGSC$, then so is $G'$.
\label{removegeneralized}
\end{lemma}

From this, we can easily obtain the following:
\begin{proposition} \label{part3}
The non-emptiness problem for $\RLG$-$\SC$ (i.e., right-linear
grammar with states and counters) is $\NP$-hard.
\end{proposition}
\begin{proof}
Again, we reduce to the subset-sum problem, which is $\NP$-hard.
An instance $I$ of this problem is the following: 

{\em Given}: Numbers $x_1, \ldots, x_k, x_{k+1}$,
represented in binary.

{\em Question}: Is there a  subset of
\{$x_1, \ldots ,x_k\}$ that sums to $x_{k+1}$?

\noindent   
From the lemma above, it is sufficient 
to construct for a given instance $I$ of the
subset-sum problem a generalized $\RLGSC$ $G$ such that
$L(G)\ne \emptyset$ if and only if 
the the answer to the instance $I$ is yes.

The construction of $G$ is straightforward.
$G$ has a terminal alphabet $\{a\}$,
nonterminals $Z, A_1, \ldots, A_{k+1}$ with $A_1$
the start nonterminal, two 1-reversal-bounded counters
$C_1$ and $C_2$, and states $q_0$ (start state), $q_1, q_f$ (accepting state).
The rules in $G$ are the following (where $v$ is either $0$ or $1$):

\begin{enumerate}
\item
$(q_0, v, 0, A_i) \rightarrow (q_0, + x_i, 0, A_{i+1})
 ~|~ (q_0, 0, 0, A_{i+1})$  for $1 \le i \le k$.  
\item
$(q_0, v, 0, A_{k+1}) \rightarrow (q_1, 0, +x_{k+1}, Z)$
\item
$(q_1, 1, 1, Z) \rightarrow (q_1, -1, -1, Z)$
\item
$(q_1, 0, 0, Z) \rightarrow (q_f, 0, 0, a)$.
\end{enumerate}

\noindent
Clearly, $L(G)$ is either empty or $\{a\}$,
and it is non-empty if and only if the instance $I$ has a solution, and it is therefore of index $1$.
By Lemma \ref{removegeneralized}, $G$ (which has only two 
1-reversal-bounded counters) can be converted
in polynomial time to an $\RLGSC$ $G'$ with 1-reversal-bounded counters, but the
number of 1-reversal-bounded counters would depend on the sizes of the
constants used in the rules of $G$.
\qed
\end{proof}

In contrast to the above results, when both the index and number
of 1-reversal-bounded counter are fixed, we have:

\begin{proposition} \label{part4}
\label{prop17}
Let  $m, k \ge 1$ be fixed.   The emptiness problem
for $m$-index $\CFGSC$ $G$ with at most $k$ 1-reversal-bounded
counters is decidable in polynomial time. Furthermore, this is also true for every
$\CFGSC$ $G$ with at most $k$ 1-reversal-bounded counters where, if $L(G) \neq \emptyset$, there
is some derivation of index $m$.
\end{proposition}
\begin{proof}
Let $G = (V,\Sigma,P,S,Q,q_0,F)$ be  an $m$-index $\CFGSC$ with at most $k$ 1-reversal-bounded
counters.  We first construct from $G$ a grammar $G'$ where all
terminal symbols are mapped to $\lambda$. Clearly, $L(G')$ is empty
if and only if $L(G)$ is empty. Then all rules in $G'$  are of the form:
$$(q,i_1, \ldots, i_k, A ) \rightarrow (p, l_1, \ldots, l_k, u),$$
where $i_j \in \{0,1\}, l_j \in \{-1,0,1\}$ for $1 \leq j \leq k$, $u \in V^*, 0 \leq |u| \leq m$.
Furthermore, since $G$ has index $m$, $L(G) \neq \emptyset$ if and only if $L(G') \neq \emptyset$
if and only if there is some derivation of $\lambda$ in $G'$ such that every sentential
form in the derivation has at most $m$ nonterminals. This is even true if there is some derivation of
index $m$ in $G$.
Now from $G'$, we construct an $\NCM$ $M$ with $k$ 1-reversal-bounded counters as follows:
Its initial state  is $[q_0, S]$.  
The other states of $M$ are of the form $[q, w]$,  where $q \in Q, w\in V^*, 0 \leq |w| \leq m$.

Then $M$ starts in state $[q_0, S]$ with all its $k$ counters zero.
A move of $M$ is defined by: if $G'$ has a rule
$$(q, i_1, \ldots, i_k, A) \rightarrow  (p, l_1, \ldots, l_k, v),$$
$i_j \in \{0,1\}, l_j \in \{-1,0,1\}$ for $1 \leq j \leq k$,
then in $M$, for all strings $xAy$ where $x,y$ are strings of nonterminals (possibly empty) with $|xAy| \leq m$ and $|xvy| \leq m$, create transitions from
state $[q, xAy]$ and counter
status $i_1, \ldots, i_k$ on $\lambda$, that go to state $[p, xvy]$ and update the
counters by  $l_1,\ldots, l_k$. Since $m$ is fixed, the number of words $xAy$ is polynomial in $|V|$, and therefore this
construction runs in polynomial time.
The accepting states of $M$ are of the form  $[p, \lambda]$, where
$p$ is an accepting state of $G'$.

Since $G'$ is of index $m$, then $L(G')$ is non-empty if and only if there is a derivation of $\lambda$ in $G'$
$$(p_0, c_{0,1},\ldots, c_{0,k},\alpha_0) \Rightarrow \cdots \Rightarrow
(p_n, c_{n,1},\ldots, c_{n,k},\alpha_n),$$
where $p_0 = q_0, \alpha_0 = S, \alpha_n = \lambda$, $p_n \in F$, and $c_{0,i} = 0, 1 \leq i \leq k,
|\alpha_j| \leq m,1 \leq j \leq n$. If this derivation exists, then there is a computation
$$([p_0,\alpha_0],\lambda, c_{0,1}, \ldots, c_{0,k}) \vdash_M \cdots \vdash_M ([p_n,\alpha_n],\lambda, c_{n,1}, \ldots, c_{n,k}),$$ where $[p_n,\alpha_n]$ is a final state. Furthermore, if there is such a computation of $M$, then there is a  corresponding derivation of $G'$ as well. Hence, $L(G') = L(M)$.
Clearly, since $G$ (and, hence, $G'$) is $m$-index and $m$ is fixed, the size of
$M$ is polynomial in the size of $G'$ (hence, of $G$).
Since $M$ is an $\NCM$ with a fixed ($k$) number of 1-reversal-bounded counters and it
is known that the emptiness problem for $\NCM$ with a fixed number of
1-reversal-bounded counters is decidable in polynomial time \cite{Gurari1981220}, the result follows.
\qed
\end{proof}

Finally, we will give an application of the results
above.  

As noted in the proof of Proposition \ref{part3}, the generalized
$\RLGSC$ $G$ constructed in that proof (which is obviously 1-index)
has two 1-reversal-bounded counters.  Although there
is a polynomial time algorithm to convert $G$ to an equivalent
$\RLGSC$ $G'$ (hence, $|G'|$ is polynomial in $|G|$),
the number of 1-reversal-bounded counters of $G'$ would 
depend on the the sizes of the length of the constants (in binary) used to increase the counters in the rules of $G$.
A question arises as to whether there is some fixed $r$, such
we can always convert in polynomial time a generalized $\RLGSC$ $G$
with two 1-reversal-bounded counters to an equivalent $\RLGSC$ $G'$ with
at most $r$ 1-reversal-bounded counters.  The following corollary says
it is unlikely.

\begin{corollary}
Let $m$ and $r$ be any fixed positive integers. If there
is a polynomial-time algorithm which can convert any
generalized $\RLG$-$\SC$ (which is obviously of index 1)
with two 1-reversal-bounded counters to an 
equivalent $\CFGSC$ with at most $r$ 1-reversal-bounded counters 
and whose index is at most $m$, then $\P$ = $\NP$.
\end{corollary}
\begin{proof}
This follows from Propositions  \ref{part3} and \ref{part4}.
\qed
\end{proof}

It is not known whether there is an infinite hierarchy of $\RLG$-$\SC$
in terms of 1-reversal-bounded counters (i.e., whether for every $k \ge 1$, there
is a $k' >  k$ such that there is a language accepted by a $\RLG$-$\SC$
with $k'$ 1- reversal-bounded counters that cannot be accepted by a
$\RLG$-$\SC$ with only $k$ 1-reversal-bounded counters). The
following corollary, which also follows from Propositions \ref{part3}
and \ref{part4}, says that an infinite hierarchy seems likely.   

\begin{corollary}
Let $k$ be any fixed positive integer.  If there is a polynomial-time
algorithm that can convert any $\RLG$-$\SC$ (which may have more than
$k$  1-reversal-bounded counters) to an  equivalent $\RLG$-$\SC$ with at most
$k$  1-reversal-bounded counters, then $\P$ = $\NP$.
\end{corollary}

\section{Conclusions and Future Directions}

We studied state grammars, and we showed that with a new circular derivation relation, they generate all recursively enumerable languages. We also studied
state grammars with stores (e.g., reversal-bounded counters) under the free interpretation and the leftmost derivation relation.
When using the free interpretation derivation relation, the counters do not add any generative capacity, and only states are needed. When using leftmost derivations, the class coincides with the machine model $\NPCM$ (pushdown automata with reversal-bounded counters). This leads to the result that state grammars with counters and leftmost derivations are strictly weaker than state grammars with no counters and the free interpretation derivation relation. 
We also investigated the complexity of the emptiness problem involving state
grammars with reversal-bounded counters. It was shown that for $m$ and $k$ fixed, the emptiness problem for $m$-index state grammars with $k$ $1$-reversal-bounded counters can be solved in polynomial time. Also, two results concerning reducing the number of counters, and allowing to add values larger than one in the   description of these grammars,  are
reduced to the question of whether $\P = \NP$.  

There are other interesting problems of descriptional complexity that are open. For example, do state grammars form an infinite hierarchy with the number of states? We conjecture that in Example \ref{CFGS}, for each $k \geq 2$, it is impossible to generate $L_k$ with a state grammar with fewer than $k$ states, which would form such a hierarchy.

\section*{Acknowledgements}
We thank the anonymous reviewers for a careful reading of the paper.

\bibliography{bounded}{}
\bibliographystyle{elsarticle-num}

\end{document}